\documentclass[english,12pt]{article}
\usepackage{amssymb}
\usepackage{babel}
\usepackage{latexsym}

\usepackage[T1]{fontenc}
\usepackage[latin9]{inputenc}
\usepackage{geometry}

\usepackage{appendix}

\usepackage{mathrsfs}
\usepackage{amssymb}
\usepackage{amsmath}
\usepackage{amsfonts, epsf,epsfig,color}
\usepackage{graphicx}
\usepackage{tensor}
\usepackage[all]{xy}
\include{macro}

\newcommand{\nn}{\nonumber}

\newcommand{\C}{\mathbb{C}}

\newcommand{\R}{\mathbb{R}}

\newcommand{\T}{\mathbb{T}}
\newcommand{\Z}{\mathbb{Z}}
\newcommand{\p}{\partial}
\newcommand{\la}{\langle}
\newcommand{\ra}{\rangle}
\newcommand{\be}{\begin{equation}\label}
\newcommand{\ee}{\end{equation}}
\newcommand{\bea}{\begin{eqnarray}\label}
\newcommand{\eea}{\end{eqnarray}}

\begin{document}

\title{Amplitudes of 3d Yang Mills Theory}
\author{\normalsize Arthur E. Lipstein\footnote{Arthur.Lipstein@maths.ox.ac.uk}$^{~1}$\& Lionel Mason\footnote{lmason@maths.ox.ac.uk}$^{~2}$ \\ \small \textit{University of Oxford} \\ \small \textit{24-29 St Giles'} \\ \small \textit{Oxford, OX1 3LB, U.K.}}
\maketitle
\begin{abstract}
This paper studies various properties of amplitudes in 3d super Yang
Mills theory. First we explain how to obtain the amplitudes of 3d super Yang
Mills theories from 4d super Yang Mills theories and obtain their helicity
structure. Next, we use a 3d BFCW recursion
relation to show that the tree amplitudes and loop integrands of
maximal 3d super Yang Mills have dual conformal covariance (although
not invariance, so that the amplitudes themselves are not dual
conformal).  Finally, we argue that the one-loop
amplitudes of maximal 3d super Yang-Mills can be reduced to scalar box
diagrams and evaluate these diagrams using dimensional
regularization. We find that the one-loop MHV amplitudes vanish and
the one-loop non-MHV amplitudes are finite.   
{\normalsize \par}
\end{abstract}

\section{Introduction}
\subsection{Background}
The study of scattering amplitudes of 4d $\mathcal{N}=4$ super
Yang-Mills (sYM) \cite{Brink:1976bc} has revealed many surprises. For
example, the amplitudes of 4d $\mathcal{N}=4$ sYM can be computed
using BCFW recursion relations (which relate higher point on-shell
amplitudes to lower-point on-shell amplitudes)
\cite{Britto:2004ap,Britto:2005fq,ArkaniHamed:2010kv} or an MHV
formalism (where Maximal Helicity Violating, MHV, amplitudes are used
as the Feynman vertices for constructing all other amplitudes)
\cite{Cachazo:2004kj,Risager:2005vk}. These techniques in turn contributed
to the discovery and proof of new symmetries and dualites. For example,
the on-shell scattering amplitudes $\mathcal{N}=4$ sYM were seen to be
dual to null polygonal Wilson loops
\cite{Alday:2007hr,Brandhuber:2007yx,Drummond:2007cf,Mason:2010yk,CaronHuot:2010ek}. This
duality, plus the superconformal symmetry of the Wilson loops implies
that the amplitudes have dual superconformal symmetry which is
inequivalent to the original superconformal symmetry
\cite{Drummond:2007aua,Brandhuber:2008pf,Drummond:2008vq}. Recently,
a Wilson loop/correlator duality has also been discovered
\cite{Alday:2010zy,Eden:2010zz,Eden:2010ce,Eden:2011yp,Eden:2011ku,Dualc,Adamo:2011dq}. The
dual superconformal symmetry of $\mathcal{N}=4$ sYM can be understood
using the $AdS/CFT$ correspondence \cite{Maldacena:1997re}. In
particular, it is related to the fact that type IIB string
theory \cite{Green:1981yb,Schwarz:1983qr} on $AdS_5 \times S^5$ is
self-dual under a certain combination of bosonic and fermionic
T-duality transformations
\cite{Berkovits:2008ic,Beisert:2008iq,Drummond:2010qh}. 

The MHV expansion of $\mathcal{N}=4$ sYM can be obtained as
the Feynamn diagrams of a
twistor space action in an axial gauge both for amplitudes
\cite{Boels:2006ir,Boels:2007qn,Adamo:2011cb} (where the
usual superconformal symmetry is manifest) or for the null polygonal
Wilson loop \cite{Bullimore:2010pj,Mason:2010yk}. The twistor action can also be used to obtain an analytic proof of the amplitude/Wilson loop duality \cite{Mason:2010yk,Bullimore:2011ni} as well as the Wilson loop correlator duality \cite{Adamo:2011dq}. Moreover, when dual superconformal symmetry is combined with ordinary superconformal symmetry, this gives Yangian symmetry, which is a hallmark of integrability \cite{Dolan:2004ps,Drummond:2009fd}. The amplitudes of $\mathcal{N}=4$ sYM can be computed using a contour integral over a Grassmannian which makes the Yangian symmetry manifest \cite{ArkaniHamed:2009dn}.

In this paper, we study the amplitudes of 3d Yang-Mills (YM)
theories. One motivation for this study is that 3d YM theories correspond to 4d YM theories at high temperature.  Although 3d YM theories can be obtained by dimensional reduction of 4d YM theories, they exhibit many properties which do not follow trivially from dimensional reduction. For example, unlike in 4d YM theory, it is possible to analytically demonstrate the existence of a mass gap in 3d YM theory using a Hamiltonian analysis \cite{Karabali:1997wk,Agarwal:2012bn}. Furthermore, 3d YM theories are intimately related to 3d Chern-Simons (CS) theories. The relationship between YM and CS theories can be motivated using string/M-theory. In particular, 3d YM theories provide a low energy description of two dimensional objects of type IIA string theory known as D2-branes \cite{Itzhaki:1998dd} and Chern-Simons theories describe two-dimensional objects of M-theory known as M2-branes \cite{Schwarz:2004yj}. Since type IIA string theory approaches M-theory at strong coupling, this implies that at strong coupling, 3d YM theories should flow to CS theories. 

More concretely, maximal 3d sYM (i.e. $\mathcal{N}=8$ sYM) with gauge group $U(N)$ should be equivalent to the ABJM theory with gauge group $U(N)_1 \times U(N)_{-1}$, where the subscript indicates the level of each $U(N)$ gauge field (the level refers to an integer which appears in the coefficient of the kinetic terms of the gauge field). The ABJM theory is a superconformal Chern-Simons theory with classical $\mathcal{N}=6$ supersymmetry and gauge group $U(N)_k \times U(N)_{-k}$ \cite{Aharony:2008ug}. When $k=1,2$, the supersymmetry is conjectured to become enhanced to $\mathcal{N}=8$. The ABJM theory is conjectured to describe a stack of $\mathcal{N}$ coincident M2-branes on a $\mathbb{Z}_k$ orbifold and to be dual to M-theory on $AdS_4 \times S^7/\mathbb{Z}_k$ with $N$ units of flux through the 7-sphere. In the limit that $N \gg k^5$, the gravity side reduces to type IIA string theory on $AdS_4 \times CP^3$. On the other hand, $\mathcal{N}=8$ sYM is conjectured describe a stack of D2 branes in flat space and to be dual to type IIA string theory on the supergravity background obtained in \cite{Itzhaki:1998dd,Lin:2005nh}. Another superconformal Chern-Simons theory which is closely related to the ABJM theory is the BLG theory, which has gauge group $SU(2)_k \times SU(2)_{-k}$ and classical $\mathcal{N}=8$ susy \cite{BLG,Bagger:2007vi,Gustavsson:2007vu}.  The BLG theory is conjectured to be equivalent to the ABJM theory with rank $N=2$ and level $k=2$ \cite{Lambert:2010ji}. Its gravity dual for general values of $k$ however, is not well-understood. The conjecture relating the ABJM theory to $\mathcal{N}=8$ sYM has been tested by matching partition functions \cite{Kapustin:2010xq} and superconformal indices \cite{Bashkirov:2011pt,Gang:2011xp}.

A great deal of progress has been made in understanding the scattering amplitudes of 3d gauge theories, especially the ABJM theory. In \cite{Gang:2010gy}, a BCFW recursion relation was developed which for 3d gauge theories and used to show that all tree amplitudes and cut-constructable loop integrands of the ABJM theory have dual superconformal symmetry, which was first proposed in \cite{Bargheer:2010hn,Huang:2010qy}. Furthermore, a matrix integral for the ABJM amplitudes was proposed in \cite{Lee:2010du}. This integral is taken over an orthogonal Grassmannian. There is also some evidence for an amplitude/Wilson loop duality in the ABJM theory. The 2-loop correction to the 4pt amplitude was computed \cite{Chen:2011vv,Bianchi:2011dg} and matched with the 2-loop correction to a bosonic 4-cusp null-polygonal Wilson loop \cite{Henn:2010ps}. Despite these encouraging results, it is not clear how to define a super-Wilson loop that is dual to higher point amplitudes in the ABJM theory. There has also been work on a Wilson-loop correlator duality in the ABJM theory \cite{Bianchi:2011rn}. Recently the 1-loop 6 and 8pt amplitudes of the ABJM theory were also computed \cite{Bianchi:2012cq,Bargheer:2012cp,Brandhuber:2012un}. 

The amplitudes of the BLG theory and $\mathcal{N}=8$ sYM have also been studied to a lesser extent. The four point tree-level superamplitude of the BLG theory was constructed in \cite{Huang:2010rn}. It was also shown that the four and six point amlitudes of the BLG theory can be squared into the amplitudes of $\mathcal{N}=16$ 3d sugra \cite{Bargheer:2012gv}. The scattering amplitudes of 3d sYM theories were studied in \cite{Chiou:2005jn,Agarwal:2011tz,Agarwal:2012jj}. In particular, \cite{Chiou:2005jn} proposed minitwistor string theories which correspond to 3d sYM theories with massive spinors and scalars. Reference \cite{Agarwal:2011tz} showed the the tree-level 4pt amplitudes of $\mathcal{N}=2,4, 8$ sYM can be expressed in such a way that they have $SO(\mathcal{N})$ symmetry and obtained the 4pt one-loop integrand of $\mathcal{N}=8$ sYM. Reference \cite{Agarwal:2012jj} extended the $SO(\mathcal{N})$ symmetry discovered in \cite{Agarwal:2011tz} to various supergroups.

\subsection{This paper}

In this paper, we study the amplitudes of 3d sYM theories from various
points of view, both by dimensional reduction with some independent checks
of the various interesting properties.  For example, despite the fact
that a YM gauge field has one polarization in 3d, we show that the
amplitudes of 3d sYM theory still have helicity structure, essentially
following from the representation theory of the $SO(7)$ $R$-symmetry
group. Furthermore, using the 3d BCFW formula proposed in
\cite{Gang:2010gy}, we prove that the tree-level amplitudes of
$\mathcal{N}=8$ sYM have dual conformal symmetry (once they are
stripped of a supermomentum delta function) and the cut-constructable
loop integrands have dual conformal covariance. In particular, under a
dual inversion, they transform like 4d loop integrands and so cannot
be integrated without breaking dual conformal invariance in 3d. It is
also the case that 3d YM does not have ordinary conformal symmetry
either because the coupling constant is dimensionful. As a result, it
is more natural to express the amplitudes in terms of minitwistors,
which break conformal invariance, rather than ordinary twistors, which
make conformal symmetry manifest. On the other hand, it is possible to
write the amplitudes in terms of 3d momentum twistors which make the
dual conformal covariance manifest (defined originally for 4d field
theories in \cite{Hodges:2009hk}). 3d momentum twistors can be
obtained by imposing constraints on 4d momentum twistors, and it
follows that when the amplitudes of $\mathcal{N}=8$ sYM are expressed
in terms of momentum twistors, they can be computed using a
Grassmannian integral formula. This formula comes from restricting the
momentum twistor Grassmannian integral formula of $\mathcal{N}=4$ sYM
\cite{Mason:2009qx,ArkaniHamed:2009vw} to the kinematics of 3d
momentum twistor space.  One can similarly compute the all-loop integrand by restricting the the momentum twistor holomorphic Wilson-loop \cite{Mason:2010yk} to 3d kinematics although we will treat these two topics in a separate paper.  

The dual inversion properties of the loop integrands of
$\mathcal{N}=8$ sYM together with direct dimensional reduction in
momentum space implies that they are the same as those of 4d
$\mathcal{N}=4$ sYM restricted to 3d. This
leads to a covariance rather than invariance of the loop integrand in
3d as the integrands have the correct weights for integration only in
4d, so that the loop amplitudes themselves are {\em not} dual conformal
invariant.  Nevertheless, since the one-loop amplitudes of $\mathcal{N}=4$ sYM can be reduced to evaluating a one-loop scalar box integral with massless propagators \cite{Bern:1994zx,Bern:2004bt,Bern:2011qt}, it follows that the same must be true for $\mathcal{N}=8$ sYM (this was verified at 4pt in \cite{Agarwal:2011tz}). Taking the loop momenta to be three-dimensional and evaluating the box integral in 3d using dimensional regularization, we find that the one-loop MHV amplitudes vanish and the one-loop non-MHV amplitudes are finite. These results are reminiscent of the loop results which have recently been obtained in the ABJM theory, where the one-loop correction to the 4pt amplitude was shown to vanish \cite{Agarwal:2008pu,Chen:2011vv}, and one-loop corrections to higher point amplitudes were found to be finite \cite{Bianchi:2012cq,Bargheer:2012cp,Brandhuber:2012un}. The fact that $\mathcal{N}=8$ sYM and the ABJM theory appear to be the only 3d field theories whose amplitudes have dual conformal covariance and that their one-loop amplitudes have a similar structure can be seen as new evidence that these two theories are related.  

The structure of this paper is as follows. In section \ref{4d23d}, we
describe how to obtain 3d sYM amplitudes (and loop integrands) by restricting 4d sYM amplitudes to 3d kinematics. In that section, we also explain how polarization and helicity can be defined in three dimensions. In section \ref{dualconf3d}, we review the 3d BCFW recursion relation and use it to show that all tree amplitudes of maximal 3d sYM have dual conformal symmetry. Unitarity then implies that the loop integrands of $\mathcal{N}=8$ sYM transform covariantly under dual inversions. We then define 3d momentum twistors which make the dual conformal covariance of $\mathcal{N}=8$ sYM manifest. In section \ref{3dloops}, we argue that the one-loop amplitudes of $\mathcal{N}=8$ sYM are given by one-loop scalar box integrals with 3d loop momenta and compute these box integrals using dimenisonal regularization. In section \ref{conclusion}, we present our conclusions and discuss possible implications of our results for 3d Chern-Simons theories. There are also several appendices which present the details of various calculations in this paper. In appendix \ref{3ptbcfw}, we compute all MHV amplitudes of $\mathcal{N}=8$ sYM using 3d BCFW. In appendix \ref{rsymm}, we discuss the R-symmetry of tree-level $\mathcal{N}=8$ sYM amplitudes. In particular, we show that the 4-point superamplitude has $SO(8)$ R-symmetry (confirming the result of \cite{Agarwal:2011tz}) and argue that higher point amplitudes exist which have $SO(7)$ R-symmetry, but not $SO(8)$ R-symmetry. In appendix \ref{dualtree}, we prove that all tree-level amplitudes of $\mathcal{N}=8$ sYM have dual conformal symmetry (once they are stripped of a supermomentum delta function). In appendix \ref{dualloop}, we use unitarity to show that the loop integrands of  $\mathcal{N}=8$  sYM have dual conformal covariance. In appendix \ref{momtwistproof}, we relate spinors of 3d momentum twistors to the spinors of the on-shell superspace. Finally, in appendix \ref{loops}, we present the calculation of one-loop scalar box integrals in three dimensions using dimensional regularization.

\section{3d Amplitudes from 4d} \label{4d23d}

The three dimensional gauge theories we will consider are obtained by
symmetry reduction of super Yang-Mills in four dimensions to
three (indeed, originally from ten). Equivalently, it can be obtained by dimensionally reducing 4d
YM and neglecting Kaluza-Klein modes. We could instead Wick rotate and
compactify the time direction in 4d. Then 3d YM would simply
correspond to 4d YM at infinitely high temperature. We could also
choose to reduce 4d YM along a spatial direction, which would give 3d
YM in Lorentzian signature.  

As far as Feynman rules are concerned, the effect of symmetry
reduction is that we have the same diagrams, vertices and propagators
as in four dimensions, but the external kinematics and the propagator
momenta (including those at all loop orders) are simply reduced to 3d.
Thus the on-shell scattering amplitudes can be obtained completely at
tree-level by restricting the external kinematics to three dimensions,
and at loop level by further restricting loop momenta to three dimensions.

With this in mind, we will now explain in more detail how to obtain on-shell
scattering amplitudes in 3d YM amplitudes from those in 4d YM. First
recall that a complex null momentum in 4d can be expressed in terms of two
spinors as follows:  
\begin{equation}
p_{\alpha \dot{\beta}}=\lambda_{\alpha}\tilde{\lambda}_{\dot{\beta}}
\label{null}
\end{equation}
where $\alpha$ and $\dot{\beta}$ are $SL(2,\C)$ indices of opposite chirality. For real null momenta in Lorentz signature, $\tilde\lambda_\alpha=\bar\lambda_\alpha$.
The two types of indices arise from the fact that the complex rotation group
in four dimensions $SO(4)$ is double covered by $SL(2,\C)_{L}\times SL(2,\C)_{R}$ reducing to $SU(2,\C)_L\times SU(2,\C)_R$ in Euclidean signature or to $SL(2,\C)$ (with the second factor being the complex conjugate of the first) in Lorentz signature .
On the other hand, the complex rotation group in three dimensions is double
covered by $SL(2,\C)$ embedded as a diagonal
subgroup $diag\left (SL(2)_{L}\times SL(2)_{R}\right)$ reducing to $SU(2)$ in Euclidean signature and $SL(2,\R)$ in Lorentzian signature. Hence, when
one reduces to three dimensions, the distinction between dotted coordinates
and undotted coordinates disappears.

Time translations in 4d are determined by a vector field
$T^{\alpha\dot\alpha}$ which we can normalize so that $T\cdot T=2$.
We can use $T$ to define the diagonal embedding of spin groups above
and to eliminate all dotted indices. Hence, we can project the 4d momentum in equation \eqref{null} by changing $\dot{\beta}$ to $\beta$ and symmetrizing the indices: 
\[
p_{\alpha\beta}=\lambda_{(\alpha}\tilde{\lambda}_{\beta)}.\]
Symmetrizing the indices projects out the time component of the 4d vector as $p_{\alpha\dot\beta}T^{\dot\beta}_\beta=p_{(\alpha\beta)}+\varepsilon_{\alpha\beta}\la\lambda\,\tilde\lambda\ra$ with the first symmetrized term giving three spatial components and the second skew term the time component. 
The magnitude of the 3-momentum is \[
p^{2}=-\frac{1}{2} p^{\alpha \beta} p_{\alpha \beta} =\frac{1}{4} \langle \lambda\tilde{\lambda}\rangle ^{2},\]
where $\langle \lambda\tilde{\lambda}\rangle= \lambda^{\alpha} \tilde{\lambda}_{\alpha}$.
If the momentum is null, this implies that $\lambda \propto \tilde{\lambda}$.  This cannot be the case for real momenta in Euclidean signature since $\la\lambda\, \hat\lambda\ra$ is positive definite (where $\hat\lambda_\alpha=(\bar\lambda_1,-\bar\lambda_0)$ is the $SU(2)$ conjugate of $\lambda$), but 
in Lorentzian signature it implies  that $\lambda=\bar \lambda$ is real. 

The on-shell amplitudes of 4d sYM theories can be parameterized using supermomenta.  These consist of $(\lambda_\alpha,\tilde\lambda_{\dot\alpha},\eta^I)$ where $(\lambda,\tilde\lambda)$ are the spinors associated with the null momentum of an external particle and $\eta^I$, $I=1,\ldots,4$ is fermionic and $I$ is an $SU(4)$ R-symmetry index (although we can easily consider lesser amounts of supersymmetry). 
 The $(\lambda_\alpha,\tilde\lambda_{\dot\alpha},\eta^I)$ are defined up to the rescaling
\be{scaling}
(\lambda_\alpha,\tilde\lambda_{\dot\alpha},\eta^I)\sim (\alpha \lambda_\alpha,\alpha^{-1}\tilde\lambda_{\dot\alpha},\alpha\eta^I)\, , \qquad \alpha\in \C^*
\ee
where we give the complex version (in Lorentz signature $\alpha\in U(1)$) and so these variables up to this equivalence give the on-shell superspace. If we wish to reduce the tree-level amplitudes of a 4d supersymmetric YM theory to 3d without breaking susy, we restrict the momenta to null vectors in 3d.  This gives 
\be{}
\tilde \lambda_{\dot\alpha}=T_{\dot\alpha}^\beta\lambda_\beta
\ee  
and this condition has the effect of fixing the rescaling freedom to just a sign.
In general we change dotted indices to undotted in the bosonic variables and leave the fermionic coordinates untouched.  Furthermore, to reduce a loop amplitude from 4d to 3d, we must similarly restrict the loop momenta in all diagrams of the 4d theory to 3d, i.e. integrate over 3d momentum space while leaving the integrand unchanged. We will discuss this further in section \ref{3dloops}. If the 4d loop amplitudes are computed using dimensional regularization with $d=4-2\epsilon$, this corresponds to letting $\epsilon = \tilde{\epsilon}+1/2$ and expanding around $\tilde{\epsilon}=0$.
  
To be more concrete, we describe the tree-level amplitudes of maximal 3d sYM as follows. We start by by dimensionally reducing 4d $\mathcal{N}=4$ sYM to three dimensions where the on-shell superfield of $\mathcal{N}=4$ sYM is
\begin{equation}
\mathcal{A}=a+\eta^{I}\psi_{I}+\frac{1}{2}\eta^{I}\eta^{J}\phi_{IJ}+\frac{1}{6}\eta^{I}\eta^{J}\eta^{K}\epsilon_{IJKL}\psi^{L}+\frac{1}{24}\eta^{I}\eta^{J}\eta^{K}\eta^{L}\epsilon_{IJKL}g\label{eq:superfield}\end{equation}
where $\phi_{IJ}=-\phi_{JI}$. Under \eqref{scaling} $\mathcal{A}$ has weight 2 so that the fields $(a,\psi_I,\phi_{IJ},\psi^I,g)$ have weights $(2,1,0,-1,-2)$ respectively; the weights  are $2h$ where $h$ is the helicity.   Thus, the fields
$a$ and $g$ contain the two on-shell degrees of freedom of a YM gauge
field, which have positive and negative helicity respectively and
the on-shell degrees of freedom consist of two helicities of a Yang-Mills field, six scalars, and
eight spinors. 

Upon dimensional reduction, the scaling symmetry is fixed to just $\pm 1$.  One of the components of the 4d gauge field becomes a scalar in three dimensions, which we shall refer to as a Higgs field. Hence, the on-shell degrees of freedom of maximal 3d sYM consist of a YM gauge field (which has one degree of freedom in 3d), seven scalars, and eight spinors.  It can be seen that there is no distinction on momentum space between a gauge field (here $a$) and a scalar, a reflection of the duality between vectors and scalars in 3-dimensions (given on space-time by $dA= *d\phi$ where $A$ is a 1-form and $\phi$ a scalar).  Thus \eqref{eq:superfield} can therefore be seen to be an on-shell
superfield of an $\mathcal{N}=8$ supermultiplet with $SO(8)$ R-symmetry expressed in the form above broken down to $SU(4)$. In this case, the fields $a$ and $g$ contain linear combinations of the Yang-Mills degree of freedom and the Higgs field.

The on-shell superspace of an $n$-pt amplitude in maximal 3d sYM is therefore given by
\[
\left\{ \lambda_{i}^{\alpha},\eta_{i}^{I}\right\} ,\,\,\, i=1,...,n
\]
where $I$ is an $SU(4)$ R-symmetry index. Given that the theory contains seven scalars, it may may at first be puzzling why only an $SU(4)$  subgroup of the $SO(7)$ R-symmetry is manifest. This is because, in order to obtain an irreducible representation, we must choose an anti-commuting subset of the basic supersymmetry generators and the maximal such subset has dimension four even with the extra supersymmetry we obtain on reduction to 3d.

In 4d, the polarization vectors of a particle with $\pm$ helicity are given by
\[
\epsilon_{+}^{\alpha\dot{\beta}}=\frac{\eta^{\alpha}\hat{\lambda}^{\dot{\beta}}}{\left\langle \eta\lambda\right\rangle },\,\,\,\epsilon_{-}^{\alpha\dot{\beta}}=\frac{\lambda^{\alpha}\hat{\eta}^{\dot{\beta}}}{\left[\hat{\lambda}\hat{\eta}\right]}=\left(\epsilon_{+}^{*}\right)^{\alpha\dot{\beta}}
\]
where $\eta^{\dot{\beta}}$ is a reference spinor. Upon reduction to three dimensions following the procedure described above, both polarization vectors reduce to 
\begin{equation}
\frac{\lambda^{(\alpha}\eta^{\beta)}}{\left\langle \lambda\eta\right\rangle }
\label{3dpol}
\end{equation}
up to a sign. Hence, there is only one type of polarization vector in three dimensions, which reflects the fact that a 3d YM field has only one degree of freedom. Hence, the helicity of the particle moving in three dimensions is not encoded in it's polarization vector. Since the gauge field has one degree of freedom in three dimensions, it can be dualized to a scalar. One can then define a $U(1)$ symmetry which rotates this scalar and the Higgs field and encodes the helicity of the particle \cite{Chiou:2005jn,Agarwal:2011tz}.  Hence, in order to define helicity in $\mathcal{N}=8$ sYM, we must break the $SO(7)$ R-symmetry to $SO(6) = SU(4)/\Z_2$, which is precisely the R-symmetry of the on-shell superspace.  

Conformal symmetry is broken by dimensional reduction (we can rescale to normalize the length of the symmetry vector) and
the coupling constant is dimensionful in three dimensions. Since the maximal 3d theory can be understood via dimensional reduction from 10 dimensions, it has $SO(7)$ $R$-symmetry. Hence the
bosonic symmetries of scattering amplitudes of maximal 3d sYM are the
Poincare group and the $SO(7)$ R-symmetry group. The full supergroup of symmetries has generators acting 
on the on-shell superspace as follows:   
$$ p^{\alpha\beta}=\lambda^\alpha\lambda^\beta\, , \qquad 
 l^\alpha\,_\beta=\lambda^\alpha\frac{\partial}{\partial \lambda^\beta}-\delta^\alpha_\beta\frac{1}{2}\lambda^\gamma\frac{\partial}{\partial \lambda^\gamma}$$
$$q^{I\alpha}=\lambda^\alpha\eta^I,\qquad q^\alpha_{I}=\lambda^\alpha\frac{\partial}{\partial \eta^I}\, , \qquad 
r_{\,\,\,\, J}^{I}=-\eta^{I}\frac{\partial}{\partial\eta^{J}}+\frac{1}{4}\delta_{J}^{I}\eta^{K}\frac{\partial}{\partial\eta^{K}}$$
where the first line gives the standard momentum space representation of the Poincar\'e group, $q^I$ and $q_I$ are the 16 $\mathcal{N}=8$ susy generators, and $r_{\,\,\,\, J}^{I}$ are the $SU(4)$ R-symmetry generators. The full $SO(8)$ R-symmetry of the on-shell superspace doesn't act linearly, but is realized by including the generators
$$
 \eta^I\eta^J\, , \qquad \frac{\p^2}{\p\eta^I\eta^J}\, , \qquad \eta^I\frac{\p}{\p\eta^I}\, . 
$$
However, we only expect the combination that lies in $SO(7)$ to act on the full S-matrix of maximal 3d sYM theory as the gauge field is distinguished  from the other bosonic fields in the nonlinear theory.  Thus the only additional generators we expect to arise as symmetries of the full theory are 
\begin{equation}
\eta^I\eta^J+\frac12  \varepsilon^{IJKL}\frac{\p^2}{\p\eta^K\eta^L}\, ,
\label{extra-gens}
\end{equation}
which can be characterized as those $SO(8)$ generators that preserve the on-shell gauge field $(1+\eta_1\eta_2\eta_3\eta_4)a$.

Using the reduction procedure we described above, it is easy to see
that the MHV amplitudes of $\mathcal{N}=8$ sYM are the same as those
of $\mathcal{N}=4$ sYM with a reduced delta function:
\[
\mathcal{A}_{n}^{MHV}=\frac{\delta^{3}\left(P\right)\delta^{8}(Q)}{\left\langle 12\right\rangle \left\langle 23\right\rangle ...\left\langle n1\right\rangle }
\]
where
$P^{\alpha\beta}=\sum_{i=1}^{n}\lambda_{i}^{\alpha}\lambda_{i}^{\beta}$
and
$Q^{I\alpha}=\sum_{i=1}^{n}\lambda_{i}^{\alpha}\eta_{i}^{I}$. However,
in three dimensions, the first nontrivial amplitude occurs at $n=4$ since all kinematic invariants vanish when $n=3$. In 4d, the vanishing of all kinematic invariants does not imply that all spinor inner products should vanish if one works in split signature. On the other hand, if all kinematic invariants vanish in 3d, this implies that all spinor inner products should vanish since $\left(p_{i}+p_{j}\right)^{2}\propto\left\langle ij\right\rangle ^{2}$. 

The reduction argument also implies that the 3d S-matrix is {\em not} invariant under the full $SO(8)$ R-symmetry of the on-shell superspace as the generator $\eta^I\p/\p\eta^I$ does not leave invariant the S-matrix for five or more particles as it has different eigenvalues on different MHV degrees, $4(k+2)$ for N$^k$MHV.  (It does however preserve the 4-particle S-matrix as there is only the MHV amplitude there.)  The extra generators \eqref{extra-gens} do however give relations between the different MHV degrees, a relation that is new to three dimensions. In appendix \ref{rsymm} we prove that the 4-point superamplitude has $SO(8)$ R-symmetry and argue that for $n>4$, there is a single $n$-point amplitude with $SO(7)$ (but not $SO(8)$) R-symmetry.  

In a future publication \cite{minitwist}, we will confirm the helicity structure of $\mathcal{N}=8$ sYM by dimensionally reducing the twistor action of maximal 4d sYM to minitwistor space.

\section{Dual Conformal Properties} \label{dualconf3d}

In this section, we will prove that the tree-level amplitudes of $\mathcal{N}=8$ sYM have dual conformal symmetry and the loop integrands have dual conformal covariance. Our approach will be similar to the one used to prove that 10d maximal sYM \cite{CaronHuot:2010rj}, 6d maximal sYM \cite{Dennen:2010dh}, 4d maximal sYM \cite{Brandhuber:2008pf}, and the ABJM theory \cite{Gang:2010gy} have dual conformal symmetry. After showing that the tree-level MHV amplitudes of $\mathcal{N}=8$ sYM have dual conformal symmetry, we use 3d BCFW to extend this to all tree-level amplitudes. Then we use unitarity to show that the cut-constructable loop integrands also have dual conformal covariance. First we review 3d BCFW recursion.

\subsection{Review of 3d BCFW} \label{bcfwreview}
Consider a 3d gauge theory with $\mathcal{N}$ amount of supersymmetry.
The amplitudes of this theory can be expressed in terms of an on-shell
superspace whose coordinates are $\left(\lambda_{i}^{\alpha},\eta_{i}^{I}\right)$,
where $\lambda$ is a bosonic, $\eta$ is fermionic, $\alpha\in\{1,2\}$,
$I\in\left\{ 1,...\mathcal{N}/2\right\} $, and $i$ labels the external
legs of the amplitude. In \cite{Gang:2010gy}, a 3d BCFW formula was proposed which involves
deforming two external legs of the amplitude in a nonlinear way. For
example, if we choose to deform legs $i$ and $j$, then the deformation
has the following form: 

\begin{equation}
\left(\begin{array}{c}
\hat{\lambda}_{i}\\
\hat{\lambda}_{j}\end{array}\right)=\left(\begin{array}{cc}
\frac{1}{2}\left(z+z^{-1}\right) & -\frac{1}{2i}\left(z-z^{-1}\right)\\
\frac{1}{2i}\left(z-z^{-1}\right) & \frac{1}{2}\left(z+z^{-1}\right)\end{array}\right)\left(\begin{array}{c}
\lambda_{i}\\
\lambda_{j}\end{array}\right)\label{eq:deform}\end{equation}
where $z$ is an arbitrary complex parameter. The same deformation
is also applied to $\left(\eta_{i},\eta_{j}\right)$. After
performing this deformation, the amplitude becomes a function of $z$.
Assuming the amplitude vanishes when $z\rightarrow\infty$, we have
the following:

\begin{equation}
\oint_{|z|=\infty}\frac{A(z)d z}{z-1}=0{\normalcolor .}\label{contour-int}
\end{equation}
On the other
hand, this contour integral must also be equal to the sum of the residues
of the integrand in the complex plane, which occur at $z=1$ and the
poles of $A(z)$. Hence, we find that

\begin{equation}
A(z=1)=-\frac{1}{2\pi i}\sum_{f,j}\int d^{\mathcal{N}/2}\eta\oint_{z_{f,j}}dz\frac{A_{L}(z,\eta)A_{R}(z,i\eta)}{\hat{p}_{f}(z)^{2}}\frac{1}{z-1}\label{eq:bcfw}\end{equation}
where $z_{f,j}$ are the poles of $A(z)$. Near its poles, $A(z)$
factorizes into two on-shell amplitudes (denoted $A_{L}$ and $A_{R}$)
multiplied by a propagator (denoted $\hat{p}_{f}(z)$). The factorization
channels are labeled by $f$. The integral $\int d^{\mathcal{N}/2}\eta$ takes into account
all the fields in the supermultiplet which can appear in the propagator. 

From the deformation in eq \eqref{eq:deform}, one can see that in any
channel, $\hat{p}_{f}(z)^{2}$ has the following form

\[
\hat{p}_{f}(z)^{2}=a_{f}z^{-2}+b_{f}+c_{f}z^{2}{\normalcolor .}\]
Hence the roots are obtained by solving a quadratic equation in $z^{2}$.
We will denote the roots $\left\{ \pm z_{1f},\pm z_{2f}\right\} $.
In terms of its roots, $\hat{p}_{f}(z)^{2}$ can be expressed as follows:

\begin{equation}
\hat{p}_{f}(z)^{2}=\frac{p_{f}^{2}\left(z^{2}-z_{1f}^{2}\right)\left(z^{2}-z_{2f}^{2}\right)}{\left(1-z_{1f}^{2}\right)\left(1-z_{2f}^{2}\right)z^{2}}
\label{phat}
\end{equation}
where $p_{f}=\hat{p}_{f}(z=1)$ is the undeformed momentum flowing
through the propagator. For factorization channels in which each subamplitude has more than three external legs one has to evaluate the residues of four simple poles, which correspond to the roots of $\hat{p}_{f}(z)^{2}$. In appendix \ref{3ptbcfw}, we explain how to evaluate factorization channels in which one or both subamplitudes has three external legs. Given the 3pt MHV superamplitude and the 3pt anti-MHV superamplitude\footnote{Although the on-shell 3pt amplitude vanishes in 3d, this only takes place on the support of the momentum conserving delta-function.  However, when the deformation parameter is such that the sum of two momenta becomes null, there is a second order pole and we need the derivative off the zero-set of the momentum conserving delta function and this gives a non-zero contribution off the form of the standard 3pt amplitude before the imposition of momentum conservation.}, we conjecture that all other superamplitudes of $\mathcal{N}=8$ sYM can be computed using 3d BCFW. We confirm this in appendix \ref{3ptbcfw} by computing all MHV superamplitudes using 3d BCFW. 



 
In deriving the 3d BCFW recursion relations, we assumed that the on-shell amplitudes vanish when the complex deformation parameter $z\rightarrow \infty$. To see that the on-shell amplitudes of $\mathcal{N}=8$ sYM have good large-$z$ behavior, consider the BCFW shift of an $n$-point MHV superamplitude:
\[
\frac{\delta^{3}(P)\delta^{8}(Q)}{\left\langle 12\right\rangle \left\langle 23\right\rangle ...\left\langle n1\right\rangle }
\]
Suppose we choose to shift legs $i$ and $j$. Note that $P$ and $Q$ are left  invariant by the supershift so the numerator of the MHV amplitude is unchanged. If the legs are adjacent, i.e. if $j=i+1$, then there is a factor of $\left\langle i-1i\right\rangle \left\langle ij\right\rangle \left\langle jj+1\right\rangle $ in the denominator. Noting that $\left\langle ij\right\rangle$ is invariant under the BCFW shift since $\left\langle ij\right\rangle ^{2} \propto \left(p_{i}+p_{j}\right)^{2}$, and that $\lambda_i(z)$ and $\lambda_j(z)$ are $\mathcal{O}(z)$ in the large-$z$ limit, we see that the MHV amplitude is $\mathcal{O}(z^{-2})$ in the large-$z$ limit. On the other hand, if the shifted legs are not adjacent, then there is a factor of $\left\langle i-1i\right\rangle \left\langle ii+1\right\rangle \left\langle jj-1\right\rangle \left\langle jj+1\right\rangle $ in the denominator so the amplitude is $\mathcal{O}(z^{-4})$ in the large-$z$ limit. This good large-$z$ behavior can be extended to non-MHV amplitudes by dimensional reduction of 4d amplitudes computed using the standard 4d MHV formalism.
In the 4d MHV formalism, all the vertices are MHV vertices, which have good large-$z$ behavior when restricted to 3d, and the propagators can only increase the fall-off in $z$, so the non-MHV amplitudes must also have good large-$z$ behavior when restricted to 3d. We have in fact obtained more fall-off than we need for the BCFW argument, so there will therefore be `bonus relations' as in gravity in addition to the recursion itself obtained by omitting the denominator in \eqref{contour-int}.  

\subsection{Tree-Level Dual Conformal Symmetry}

In this subsection, we will demonstrate that the tree-level amplitudes
of maximal 3d sYM have dual conformal symmetry. This symmetry can
be seen by expressing the amplitudes in terms of `region momenta' or
`dual' coordinates, which are defined as follows:
\begin{equation}
\left(x_{i}-x_{i+1}\right)^{\alpha\beta}=\lambda_{i}^{\alpha}\lambda_{i}^{\beta},\,\,\,\left(\theta_{i}-\theta_{i+1}\right)^{I\alpha}=\lambda_{i}^{\alpha}\eta_{i}^{I}{\normalcolor .}\label{eq:hyper}\end{equation}
The dual coordinates $x_{i}$ are the vertices of
a polygon whose edges correspond to the external momenta of the amplitude,
and the dual coordinates $\theta_{i}$ correspond to the vertices
of a polygon whose edges correspond to the external supermomenta.
In the dual space, one can then define the following inversions: 
\begin{equation}
I\left[x_{i}^{\alpha\beta}\right]=\frac{x_{i}^{\alpha\beta}}{x_{i}^{2}},\,\,\, I\left[\theta_{i}^{I\alpha}\right]=\frac{x_{i}^{\alpha\beta}\theta_{i\beta}^I}{x_{i}^{2}}{\normalcolor .}\label{eq:inverse}\end{equation}
The statement of dual conformal symmetry is then equivalent to saying
that when the amplitudes are expressed in terms of dual coordinates,
they transform covariantly under dual inversions. 

First we will demonstrate that the MHV amplitudes have dual conformal
symmetry. Then we will use the 3d BCFW recursion relation to show
that all tree-level amplitudes have this property, since they can
be constructed from MHV amplitudes. Note
that the supermomentum delta functions can be written in dual coordinates
as follows: $\delta^{3}(P)=\delta^{3}\left(x_{1}-x_{n+1}\right)$,
$\delta^{8}(Q)=\delta^{8}\left(\theta_{1}-\theta_{n+1}\right)$. Using
eqs \eqref{eq:inverse}, it follows that 

\begin{equation}
I\left[\delta^{3}\left(x_{1}-x_{n+1}\right)\right]=x_{1}^{6}\delta^{3}\left(x_{1}-x_{n+1}\right),\,\,\, I\left[\delta^{8}\left(\theta_{1}-\theta_{n+1}\right)\right]=x_{1}^{-8}\delta^{8}\left(\theta_{1}-\theta_{n+1}\right){\normalcolor .}\label{eq:delta}\end{equation}
On the other hand, the spinor inner products which appear in the Park-Taylor
formula can be written as follows: $\left\langle ii+1\right\rangle ^{2}=p_{i}\cdot p_{i+1}=\frac{1}{2}\left(p_{i}+p_{i+1}\right)^{2}=\frac{1}{2}\left(x_{i}-x_{i+2}\right)^{2}$.
Noting that $I\left[\left(x_{i}-x_{j}\right)^{2}\right]=\frac{\left(x_{i}-x_{j}\right)^{2}}{x_{i}^{2}x_{j}^{2}}$,
we therefore find that\begin{equation}
I\left[\left\langle ii+1\right\rangle \right]=\frac{\left\langle ii+1\right\rangle }{\sqrt{x_{i}^{2}x_{i+1}^{2}}}{\normalcolor .}\label{eq:inner}\end{equation}
Using eqs \eqref{eq:delta} and \eqref{eq:inner}, we therefore find that 

\[
I\left[\frac{\delta^{3}(P)\delta^{8}(Q)}{\left\langle 12\right\rangle \left\langle 23\right\rangle ...\left\langle n1\right\rangle }\right]=\frac{1}{x_{1}^{2}}\left(x_{1}^{2}x_{2}^{2}...x_{n}^{2}\right)\frac{\delta^{3}(P)\delta^{8}(Q)}{\left\langle 12\right\rangle \left\langle 23\right\rangle ...\left\langle n1\right\rangle }{\normalcolor .}\]
Hence, the MHV amplitude almost transforms covariantly under a dual
inversion. The term which spoils the covariance is the $1/x_{1}^{2}$
term in the front, which comes from the dual inversion of the supermomentum
delta function $\delta^{3}(P)\delta^{8}(Q)$. Since all tree-level
amplitudes can be constructed from MHV amplitudes, this suggests that
if we take a general tree-level amplitude and strip off
the supermomentum delta function, the remaining object will transform
covariantly under dual inversions. In other words, if we decompose
an $n$-point amplitude as follows \begin{equation}
A_{n}=\delta^{3}(P)\delta^{8}(Q)f_{n}{\normalcolor ,}\label{eq:decomp}\end{equation}
then

\begin{equation}
I\left[f_{n}\right]=x_{1}^{2}x_{2}^{2}...x_{n}^{2}f_{n}\label{eq:state}\end{equation}
We will prove this in appendix \ref{dualtree} using the 3d BCFW recursion relation described
in the previous subsection.

\subsection{Loop-Level Dual Conformal Covariance}

In appendix \ref{dualloop} we will show that the cut-constructable integrand of a loop diagram in $\mathcal{N}=8$ sYM with $n$ external legs and $L$ loops must transform as follows under a dual inversion:
\begin{equation}
I\left[\mathcal{I}_{n}^{L}\right]=\Pi_{i=1}^{n}x_{i}^{2}\Pi_{j=1}^{L}\left(x_{j}^{2}\right)^{4}\mathcal{I}_{n}^{L},
\label{integrandinversion1}
\end{equation}
where $i$ runs over the external regions and $j$ runs over loop regions in the dual space.  For example,  in Fig \ref{dualloopdiag}, regions 1 through 4 are external and regions 5 and 6 are loop regions. Eq \eqref{integrandinversion1} follows from unitarity and the dual conformal symmetry of tree-level amplitudes demonstrated in the previous section.   

In $d$ dimensions, the loop integration measure will give an additional factor of $\Pi_{j=1}^{L}\left(x_{j}^{2}\right)^{-d}$ under dual inversion. Hence, the $\Pi_{j=1}^{L}\left(x_{j}^{2}\right)^{4}$ term appearing in eq \eqref{integrandinversion1} will only be canceled out by the dual inversion of a four dimensional integral measure. Hence, we find that the loop amplitudes of maximal 3d sYM have dual conformal symmetry if the loop momenta are allowed to be four dimensional, but not if they are three dimensional. This is similar to maximal 6d sYM, whose loop amplitudes have dual conformal symmetry if the loop momenta are restricted to be four dimensional, but not if they are six dimensional. Even if the dual inversion weight of the measure balanced that of the integrand, dual conformal symmetry would still generally be broken when we evaluate the integral because we would introduce a regulator.  

\subsection{3d Momentum Twistors}

Using 3d momentum twistors, it is possible to express the amplitudes
of $\mathcal{N}=8$ sYM in a way that makes the dual conformal covariance
manifest. In 4d, the $x$ coordinates of the dual space are related to the 4d momentum twistor coordinates
as follows:
\[
x_i^{\alpha\dot{\beta}}=\frac{\lambda_{i}^{\alpha}\mu_{i-1}^{\dot{\beta}}-\lambda_{i-1}^{\alpha}\mu_{i}^{\dot{\beta}}}{\left\langle ii-1\right\rangle }\]
where $\left\langle ii-1\right\rangle =\lambda_{i}^{\alpha}\lambda_{i-1\alpha}$.
To restrict this formula to 3d, first use the time-like vector $T^{\alpha\dot{\alpha}}$
to change dotted indices into undotted indices:
\begin{equation}
x_i^{\alpha\beta}=\frac{\tilde{\lambda}_{i}^{\alpha}\tilde{\mu}_{i-1}^{\beta}-\tilde{\lambda}_{i-1}^{\alpha}\tilde{\mu}_{i}^{\beta}}{\left\langle \tilde{i}\tilde{i}-1\right\rangle },\label{eq:3dx}\end{equation}
where we have added tildes to indicate that these are 3d momentum
twistor coordinates rather than 4d momentum twistor coordinates. Since
$x^{\alpha\beta}$ should be symmetric in three dimensions, this implies
that $\tilde{\lambda}_{i}^{[\alpha}\tilde{\mu}_{i-1}^{\beta]}-\tilde{\lambda}_{i-1}^{[\alpha}\tilde{\mu}_{i}^{\beta]}=\epsilon^{\alpha\beta}\left(\tilde{\lambda}_{i}\cdot\tilde{\mu}_{i-1}-\tilde{\lambda}_{i-1}\cdot\tilde{\mu}_{i}\right)=0$,
where $\tilde{\lambda}\cdot\tilde{\mu}=\tilde{\lambda}^{\alpha} \tilde{\mu}_{\alpha}$. 
Hence, the restriction to three dimensions gives the following constraint
on the bosonic momentum twistors \cite{Lipstein:2011ej} \footnote{Equations \ref{eq:3dconstraint} and \ref{eq:lambdaprop} were obtained in collaboration with Dongmin Gang, Yu-tin Huang, Eunkyung Koh, and Sangmin Lee.}:
\begin{equation}
\left[\tilde{i}\tilde{i}-1\right]=0{\normalcolor ,}\label{eq:3dconstraint}\end{equation}
where we have defined
\begin{equation}
\left[\tilde{i}\tilde{j}\right]\equiv\tilde{\lambda}_{i}\cdot\tilde{\mu}_{j}-\tilde{\lambda}_{j}\cdot\tilde{\mu}_{i}{\normalcolor .}\label{eq:bracket}\end{equation}
Note that this bracket is manifestly dual conformal invariant. To
see this, note that $\left[\tilde{i}\tilde{j}\right]=\Omega_{AB}W_{i}^{A}W_{j}^{B}$,
where $W_{i}^{A}=\left(\tilde{\lambda}_{i}^{\alpha},\tilde{\mu}_{i\beta}\right)$
is the bosonic part of the 3d momentum twistor, and $\Omega_{AB}=\left(\begin{array}{cc}
0 & 1\\
-1 & 0\end{array}\right)$ is an invariant tensor of the dual conformal group $Sp(4)$.

Note that the spinor $\tilde{\lambda}_{i}^{\alpha}$ which appears
in the 3d momentum twistor is not the same as the spinor $\lambda_{i}^{\alpha}$
which appears in the on-shell superspace. In appendix \ref{momtwistproof}, we will show that they are proportional
to eachother and are related as follows:\begin{equation}
\lambda_{i}^{\alpha}=\sqrt{\frac{\left[\tilde{i}-1\tilde{i}+1\right]}{\left\langle \tilde{i}\tilde{i}-1\right\rangle \left\langle \tilde{i}+1\tilde{i}\right\rangle }}\tilde{\lambda}_{i}^{\alpha}{\normalcolor .}\label{eq:lambdaprop}\end{equation}
Note that the scale of the $\lambda$'s are fixed by eq \ref{eq:hyper}. On the other hand, we are free to rescale the $\tilde{\lambda}$ since the $\lambda$'s are invariant under rescalings of $\tilde{\lambda}$, as can be seen from the above equation. This is consistent with the fact that the $\tilde{\lambda}$ are spinors of 3d momentum twistors, which are projective coordinates.   

\section{Loop Amplitudes} \label{3dloops}

In section \ref{dualconf3d}, we used the 3d BCFW recursion relation to show that all tree-level amplitudes of $\mathcal{N}=8$ sYM have dual conformal symmetry. Using unitarity, we then found that all cut-constructable loop integrands transform like 4d loop integrands under a dual inversion, i.e. the loop amplitudes have dual conformal symmetry if the loop momenta are taken to be four dimensional. This is consistent with the intuition that the loop amplitudes of 3d YM can be obtained by restricting 4d YM amplitudes to 3d kinematics. In particular, to obtain a loop amplitude in 3d YM, one simply takes the corresponding 4d loop integrand and integrates over three-dimensional loop momenta.

\begin{figure}
\begin{center}
\includegraphics[scale=0.2]{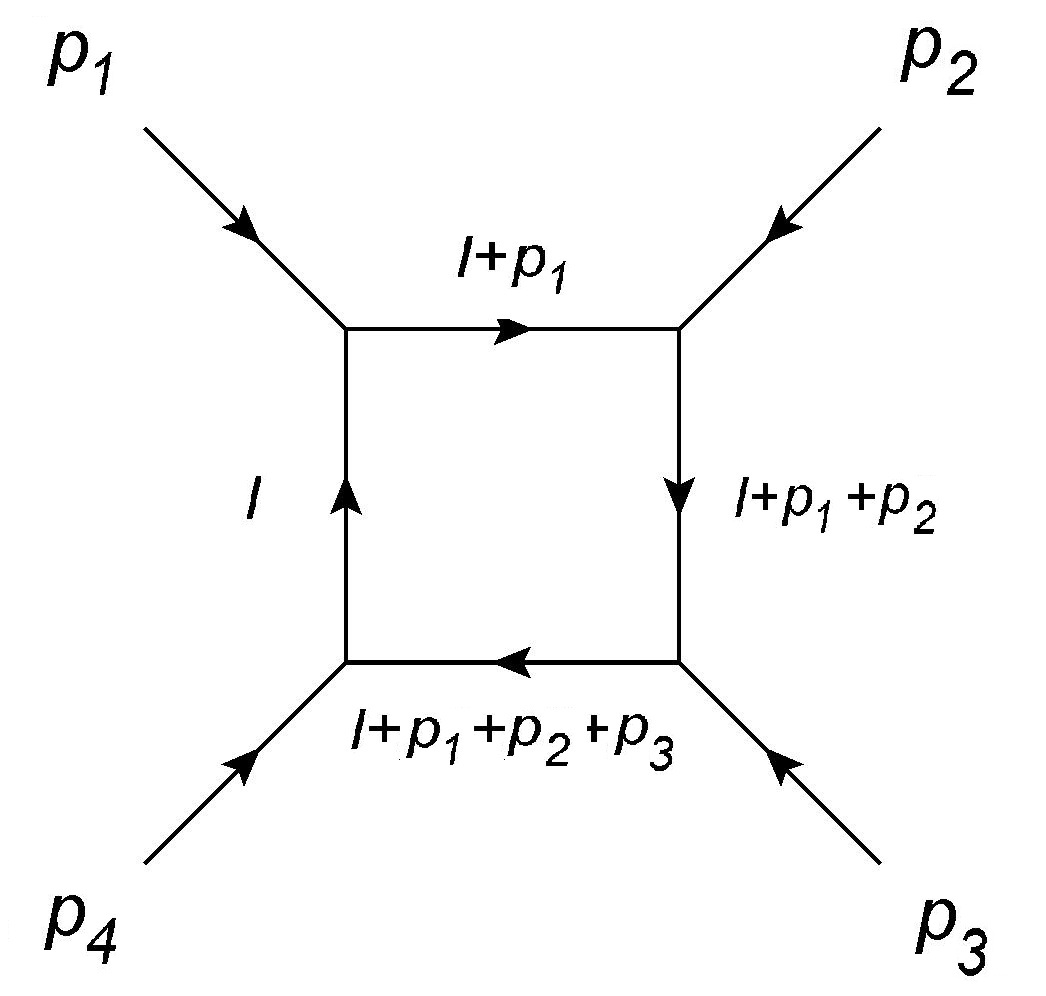}
\caption{One-loop scalar box diagram.}
\label{boxfig}
\end{center}
\end{figure}

Hence, we can obtain the one-loop amplitudes of $\mathcal{N}=8$ sYM using one-loop results from 4d $\mathcal{N}=4$ sYM. In particular, the one-loop amplitudes of $\mathcal{N}=4$ sYM can be reduced to a scalar box diagram, which is depicted in Fig \ref{boxfig}. In $d$ dimensions, this diagram is given by the following integral:
\begin{multline}
I_{4}\left(p_{1},p_{2},p_{3},p_{4}\right)=\frac{\left(\mu^{2}\right)^{2-d/2}}{\left(2\pi\right)^{d}}
\times \\
{\int\frac{d^{d}l}
{\left(l^{2}+i\epsilon\right)\left(\left(l+p_{1}\right)^{2}+i\epsilon\right)\left(\left(l+p_{1}+p_{2}\right)^{2}+i\epsilon\right)\left(\left(l+p_{1}+p_{2}+p_{2}\right)^{2}+i\epsilon\right)}}
\end{multline}
where $\mu$ is a renormalization scale and $\epsilon$ is an infinitesimal real parameter corresponding to the Feynman prescription for time-ordering. Since the integral has four scalar propagators, it will be UV finite in $d<8$ spacetime dimensions. On the other hand, it is generally IR divergent in $d\leq4$ dimensions. These divergences can be regulated using dimensional regularization. Although the propagators in the box diagram are massless, we allow the external legs to have masses:
\[
p_{i}^{2}=m_{i}^{2},\,\,\, i=1,..,4.
\] 
We also define the Mandelstam variables $s=\left(p_{1}+p_{2}\right)^{2}$, $t=\left(p_{2}+p_{3}\right)^{2}$. It is useful to distinguish between the cases where all external legs are massive, three external legs are massive, two adjacent legs are massive, two non-adjacent legs are massive, one leg is massive, and no legs are massive:  
\[
I_{4}^{4m}=I_{4}\left(s,t;m_{1}^{2},m_{2}^{2},m_{3}^{2},m_{4}^{2}\right),\,\,\, I_{4}^{3m}=I_{4}\left(s,t;0,m_{2}^{2},m_{3}^{2},m_{4}^{2}\right),I_{4}^{2mh}=I_{4}\left(s,t;0,0,m_{3}^{2},m_{4}^{2}\right)
\]
\[
I_{4}^{2me}=I_{4}\left(s,t;0,m_{2}^{2},0,m_{4}^{2}\right),\,\,\, I_{4}^{1m}=I_{4}\left(s,t;0,0,0,m_{4}^{2}\right),I_{4}^{0m}=I_{4}\left(s,t;0,0,0,0\right).
\]
These cases are referred to as the 4-mass box integral, the 3-mass box integral, the 2-mass "hard" box integral, the 2-mass "easy" box integral, the 1-mass box integral, and the massless box integral.

When all four external legs are massive,
the integral is IR finite. The four mass box integral can therefore be evaluated in $d<8$ without using dimensional regularization.
The calculation of $I^{4m}_4$ in four dimensions was carried out in \cite{'tHooft:1978xw,Denner:1991qq,Duplancic:2002dh,Hodges:2010kq}. We perform the analogous calculation in 3d following the approach of \cite{Duplancic:2002dh}, which evaluated $I_{4}^{4m}$ in 4d using Feynman parameters. Our final result for $I^{4m}_4$ is rather complicated and is presented in the attached Mathematica notebook. More details can be found in appendix \ref{loops}. 

In \cite{Bern:1992em,Bern:1993kr}, the integrals $I_{4}^{3m}, I_{4}^{2mh}, I_{4}^{2me}, I_{4}^{1m}$, and $I_{4}^{0m}$ were evaluated in four dimensions using a partial differential equation technique. They were also evaluated directly using Feynman parameters in \cite{Duplancic:2000sk}. In this paper, we will follow the approach of \cite{Duplancic:2000sk}. Setting $d=3+2\tilde{\epsilon}$ and expanding about $\tilde{\epsilon}=0$, we find that the only nonvanishing integrals are $I_{4}^{3m}$ and $I_{4}^{2mh}$, which are given by    
\[
I_{4}^{3m}=\frac{i\Gamma\left(\frac{3}{2}\right)}{(4\pi)^{2}\left(4\pi\mu^{2}\right)^{-\frac{1}{2}}}\frac{1}{st-m_{2}^{2}m_{4}^{2}}\left(R\left(t,m_{2}^{2},m_{3}^{3}\right)+R\left(s,m_{4}^{2},m_{3}^{3}\right)\right)
\]
and
\[
I_{4}^{2mh}=\frac{i\Gamma\left(\frac{3}{2}\right)}{(4\pi)^{2}\left(4\pi\mu^{2}\right)^{-\frac{1}{2}}}\frac{1}{st}\left(R\left(t,0,m_{3}^{3}\right)+R\left(s,m_{4}^{2},m_{3}^{3}\right)\right)
\]
where
\[
R\left(\alpha,\beta,m_{3}^{2}\right)=2\pi\left(m_{3}^{2}-i\epsilon\right)^{-1/2}\left(\frac{z}{\sqrt{1-z}}+\left(\frac{m_{3}^{2}+i\epsilon}{\alpha+i\epsilon}\right)\frac{1}{z}\left(2-\frac{2-z}{\sqrt{1-z}}\right)\right)
\]
\[
R\left(\alpha,0,m_{3}^{3}\right)=4\pi\left(m_{3}^{2}-i\epsilon\right)^{-1/2}\left(\frac{m_{3}^{2}+i\epsilon}{\alpha+i\epsilon}\right)
\]
and $z=1-\frac{\beta+i\epsilon}{\alpha+i\epsilon}$. Note that $R\left(\alpha,0,m_{3}^{3}\right)$ is not obtained by setting $\beta=0$ in $R\left(\alpha,\beta,m_{3}^{2}\right)$. The calculational details are described in appendix \ref{loops}. 

The one-loop MHV amplitudes of 4d $\mathcal{N}=4$ sYM can be reduced to a sum of two-mass easy diagrams \cite{Bern:1994zx} and the one-loop non-MHV amplitudes can be reduced to four-mass, three-mass and two-mass hard diagrams. It follows that using dimensional regularization, the one-loop MHV amplitudes of $\mathcal{N}=8$ sYM vanish and the one-loop non-MHV amplitudes are finite. We expect that IR divergences in $\mathcal{N}=8$ sYM will appear at two-loops, for reasons which we describe in the next paragraph.

Note that the one-loop correction to the 4pt amplitude of the ABJM theory vanishes \cite{Agarwal:2008pu,Chen:2011vv} and the one-loop correction to the six and eight-point amplitudes are finite \cite{Bianchi:2012cq,Bargheer:2012cp,Brandhuber:2012un}. Since the amplitudes of the ABJM theory are analogous to the helicity-conserving amplitudes of $\mathcal{N}=8$ sYM, perhaps it is not so surprising that the one-loop MHV amplitudes of $\mathcal{N}=8$ sYM vanish and the one-loop non-MHV amplitudes are finite. It was also shown that the two-loop 4pt ABJM amplitude has a very similar structure to the one-loop 4pt $\mathcal{N}=4$ sYM amplitude \cite{Chen:2011vv,Bianchi:2011dg,Henn:2010ps}. This suggests that the 2-loop 4pt MHV amplitude of $\mathcal{N}=8$ sYM is similar to the one-loop 4pt MHV amplitude of $\mathcal{N}=4$ sYM. It would be interesting to check this.

\section{Conclusion} \label{conclusion}

In this paper, we study color-ordered scattering amplitudes of 3d super Yang-Mills theories in the planar limit. First we describe how to obtain 3d sYM amplitudes from 4d sYM amplitudes and argue that 3d sYM amplitudes have helicity structure, even though a gluon has only one polarization in 3d. In a future publication \cite{minitwist}, we will see this for $\mathcal{N}=8$ sYM also by dimensionally reducing the twistor action of 4d $\mathcal{N}=4$ sYM to minitwistor space. 

In order to define helicity in $\mathcal{N}=8$ sYM, we must break the $SO(7)$ symmetry to $SU(4)$. Nevertheless, $SO(7)$ R-symmetry is realized nonlinearly on the amplitudes. In particular, we saw that the 4-point superamplitude has $SO(8)$ R-symmetry, confirming the results of \cite{Agarwal:2011tz}. For $n>4$, the amplitudes do not have $SO(8)$ R-symmetry; the $SO(7)$ R-symmetry mixes amplitudes of different MHV degree; it is likely that the  $n$-point amplitude is determined from its MHV and anti-MHV sector by the $SO(7)$ R-symmetry. It would be interesting to explicitly exploit the $SO(7)$ R-symmetry for more than four external legs.

Using the 3d BCFW recursion relations, we show that the tree-level amplitudes of $\mathcal{N}=8$ sYM have dual conformal symmetry if they are stripped of a supermomentum delta function. We then use unitarity to show that the cut-constructable loop integrands of $\mathcal{N}=8$ sYM transform like 4d loop integrands under dual inversions. We then defined 3d momentum twistors to parametrize the external momenta, which make the dual conformal covariance of $\mathcal{N}=8$ sYM amplitudes manifest. In terms of these variables, it is possible to compute $\mathcal{N}=8$ sYM amplitudes by restricting the Grassmannian integral formula of $\mathcal{N}=4$ sYM \cite{Mason:2009qx,ArkaniHamed:2009vw} or the 4-dimensional holomorphic Wilson loop in twistor space \cite{Mason:2010yk} to 3d momentum twistor space. Note that the dual superconformal symmetry of $\mathcal{N}=4$ sYM amplitudes follows from the fact that type IIB string theory on $AdS_5 \times S^5$ is self-dual under certain combinations of bosonic and fermionic T-dualities \cite{Berkovits:2008ic,Beisert:2008iq}. It would be interesting to see if the dual conformal covariance of $\mathcal{N}=8$ sYM amplitudes is related to T-duality of the string theory dual to $\mathcal{N}=8$ sYM, which was obtained in \cite{Itzhaki:1998dd,Lin:2005nh}.         

Since the amplitudes of 3d YM can be obtained by restricting 4d YM amplitudes to 3d kinematics, this implies that it should be possible to compute a loop amplitude in 3d YM by taking the loop integrand of the corresponding 4d loop amplitude and integrating over 3d loop momenta. For $\mathcal{N}=8$ sYM, this argument is confirmed by the fact that the loop integrands of $\mathcal{N}=8$ sYM transform like 4d $\mathcal{N}=4$ sYM loop integrands under dual conformal transformations. Since the one-loop amplitudes of $\mathcal{N}=4$ sYM can be reduced to scalar one-loop box diagrams, the same must also be true for $\mathcal{N}=8$ sYM. Evaluating these box diagrams in three dimensions using dimensional regularization, we find that the one-loop corrections to the MHV amplitudes of $\mathcal{N}=8$ sYM vanish and the one-loop corrections to non-MHV amplitudes are finite.

Although the study of 3d YM amplitudes is interesting in its own right, it also provides new insights into 3d Chern-Simons theories. Since sYM theories describe D2-branes and superconformal Chern-Simons theories describe M2-branes, string theory considerations suggest that 3d sYM theories should flow to superconformal Chern-Simons theories. For example, there is evidence that $\mathcal{N}=8$ sYM with gauge group $U(N)$ flows to the ABJM theory with gauge group $U(N)_1 \times U(N)_{-1}$. Our finding that the amplitudes $\mathcal{N}=8$ sYM have dual conformal covariance provides further evidence that it is related to the ABJM theory.

It would be very interesting to understand how the amplitudes of $\mathcal{N}=8$ sYM are related to those of the ABJM theory. It has recently been shown that the one-loop 4pt amplitude of the ABJM theory vanishes and the one-loop 6 and 8pt amplitudes are finite. On the other hand, we have found that that the one-loop MHV amplitudes of $\mathcal{N}=8$ sYM vanish and the one-loop non-MHV amplitudes are finite if one uses dimensional regularization. It would be interesting to compute two-loop amplitudes in $\mathcal{N}=8$ sYM. Since IR diveregences appear at two loops in ABJM theory, we expect them to appear at two loops in $\mathcal{N}=8$ sYM as well. It would also be interesting to compute loop amplitudes using massive regularization. 

It would also be interesting to understand how the amplitudes of $\mathcal{N}=8$ sYM are related to those of the BLG theory, which corresponds to the ABJM theory with gauge group $SU(2)_k \times SU(2)_{-k}$. In fact, the amplitudes of $\mathcal{N}=8$ sYM may be more directly related to those of the BLG theory than the ABJM theory since the on-shell superamplitudes of the BLG theory have manifest $SU(4)$ R-symmetry and  nonlinearly realized $SO(8)$ R-symmetry \cite{Huang:2010rn}. Furthermore, it was shown that when the BLG theory is given a vev, it reduces to $\mathcal{N}=8$ sYM plus terms which vanish as the vev becomes infinitely large \cite{Mukhi:2008ux}. 

In summary, although the amplitudes of 3d Yang-Mills theories can be obtained by restricting 4d Yang-Mills amplitudes to 3d kinematics, the amplitudes of 3d YM theories exhibit many interesting properties which do not follow trivially from this fact. Furthermore, we believe that the the study of scattering amplitudes of 3d YM theories will lead to a deeper understanding of 3d Chern-Simons theories.     

\section*{Acknowledgments}

We are grateful to Mathew Bullimore, James Drummond, Yu-tin Huang, and Gabriele Travaglini for helpful discussions. We would also like to thank Dongmin Gang, Yu-tin Huang, Eunkyung Koh, and Sangmin Lee for collaborating on the definition of 3d momentum twistors. AL is supported by a Simons Postdoctoral Fellowship; LM is supported by a Leverhulme Fellowship. Part of this work was completed at the "Recent Advances in Scattering Amplitudes" workshop at the Isaac Newton Institute for Mathematical Sciences in Cambridge. 

\newpage


\appendix

\section{MHV Amplitudes from 3d BCFW \label{3ptbcfw}}
In this appendix we will compute all the MHV superamplitudes of $\mathcal{N}=8$ sYM using 3d BCFW. The basic building blocks are the 3pt MHV superamplitude and the 3pt anti-MHV superamplitudes stripped of the bosonic part of their momentum conserving delta-functions.  For this reason we will strip off the momentum conservation delta-function throughout this appendix.  They are given by
\[
A_{3}^{MHV}=\frac{\delta^{8}(Q)}{\left\langle 12\right\rangle \left\langle 23\right\rangle \left\langle 31\right\rangle },\,\,\, A_{3}^{\bar{MHV}}=\frac{\delta^{4}\left(\eta_{1}\left\langle 23\right\rangle +\eta_{2}\left\langle 31\right\rangle +\eta_{3}\left\langle 12\right\rangle \right)}{\left\langle 12\right\rangle \left\langle 23\right\rangle \left\langle 31\right\rangle }
\]
where  $Q=\sum_{i=1}^{3}\lambda_{i}\eta_{i}$. These vanish when momentum conservation $P:=\sum_{i=1}^{3}\lambda_{i}\lambda_{i}=0$ is imposed as taking $p_3$ over the other side of the equation and squaring implies that $\la 1\, 2\ra^2=0$ and similarly for the other inner products and so the spinors are all proportional.  The denominator therefore vanishes to 3rd order in $P$, but the numerator to 4th order so overall the amplitude vanishes.  We will see, however, that we need the derivative of these terms off the zero of $P$ so that they do not give a nontrivial contribution and indeed generate the general amplitude.

\begin{figure}
\begin{center}
\includegraphics[scale=0.2]{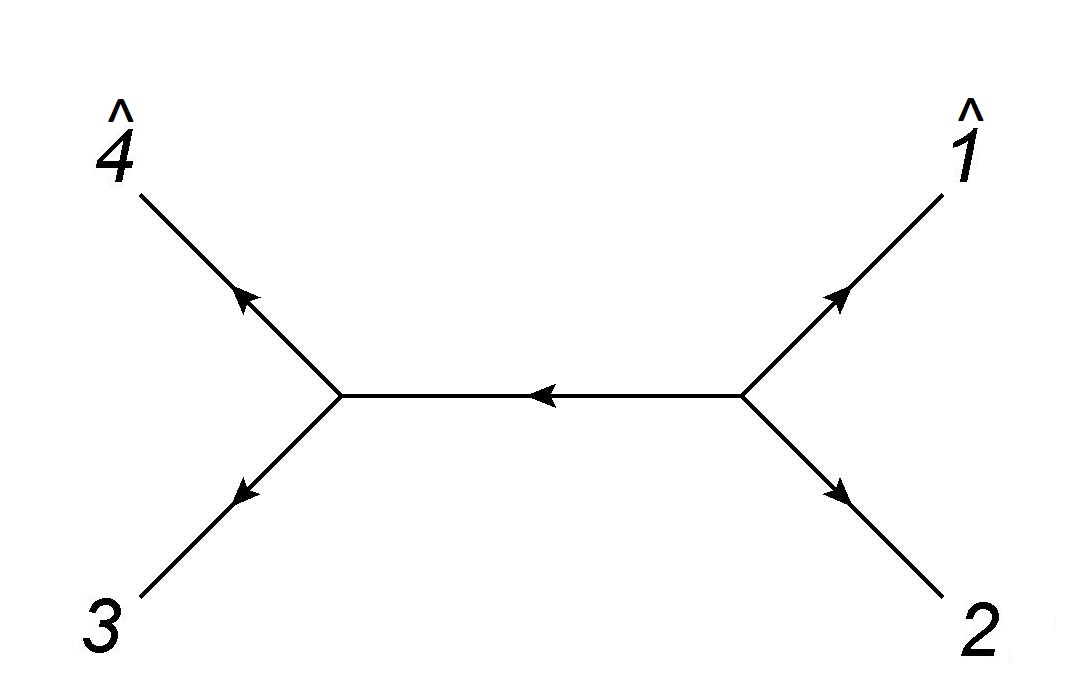}
\caption{BCFW diagram for the 4pt superamplitude.}
\label{4ptbcfw}
\end{center}
\end{figure}
Let's use these building blocks to compute the 4pt MHV superamplitude.
If we deform legs 1 and 4 according to eq \eqref{eq:deform}, then the amplitude will
become a function of the complex BCFW parameter $z$, and near the
poles in $z$, the amplitude will factorize into an on-shell 3pt MHV
superamplitude times an on-shell 3pt anti-MHV superamplitude connected
by a propagator. This is depicted in Figure \ref{4ptbcfw}. If we take the amplitude
on the left to be an MHV amplitude and the amplitude on the right to
be an anti-MHV amplitude, then near the poles we have
\[
\frac{A_{L}A_{R}}{\hat{p}^{2}}=\frac{1}{\left\langle 3\hat{4}\right\rangle \left\langle \hat{4}-\hat{p}\right\rangle \left\langle -\hat{p}3\right\rangle }\frac{1}{\left\langle 2\hat{p}\right\rangle \left\langle \hat{p}\hat{1}\right\rangle \left\langle \hat{1}2\right\rangle }\frac{\delta^{4}\left(Q_{R}\right)}{\left\langle \hat{1}2\right\rangle ^{2}}
\]
where we have left out an overall supermomentum delta function, $\hat{p}$ is the momentum flowing through the propagator, and $\delta^{4}\left(Q_{R}\right)=\delta^{4}\left(\hat{\eta}_{1}\left\langle 2\hat{p}\right\rangle +\eta_{2}\left\langle \hat{p}\hat{1}\right\rangle +\eta_{\hat{p}}\left\langle \hat{1}2\right\rangle \right)$. Note that $\left\langle \hat{4}-\hat{p}\right\rangle \left\langle 2\hat{p}\right\rangle =i\left\langle 41\right\rangle \left\langle \hat{1}2\right\rangle $
and $\left\langle -\hat{p}3\right\rangle \left\langle \hat{p}\hat{1}\right\rangle =i\left\langle \hat{1}2\right\rangle \left\langle 23\right\rangle $.
Furthermore, if we integrate $\delta^{4}\left(Q_{R}\right)$ over $\eta_{\hat{p}}$, we get a factor of $\left\langle \hat{1}2\right\rangle ^{4}$.
Hence, we find that 
\begin{equation}
\int d^{4}\eta_{\hat{p}}\frac{A_{L}A_{R}}{\hat{p}^{2}}=\frac{\delta^{8}(Q)}{\left\langle 41\right\rangle \left\langle 23\right\rangle \left\langle 3\hat{4}\right\rangle \left\langle \hat{1}2\right\rangle }=\frac{\delta^{8}(Q)}{\left\langle 41\right\rangle \left\langle 23\right\rangle \left\langle \hat{1}2\right\rangle ^{2}}\label{eq:mhv1}
\end{equation}
where noted that $\left\langle \hat{1}2\right\rangle =\left\langle 3\hat{4}\right\rangle $
for 3d four-point kinematics. From eq \eqref{eq:deform}, we see that 
\begin{equation}
\left\langle \hat{1}2\right\rangle =\frac{1}{2}\left(z+z^{-1}\right)\left\langle 12\right\rangle +\frac{i}{2}\left(z-z^{-1}\right)\left\langle 42\right\rangle =\frac{\left\langle 12\right\rangle \left(z^{2}-z_{0}^{2}\right)}{z\left(1-z_{0}^{2}\right)},\,\,\, z_{0}^{2}=-\frac{\left\langle 12\right\rangle -i\left\langle 42\right\rangle }{\left\langle 12\right\rangle +i\left\langle 42\right\rangle }\label{eq:mhv2}
\end{equation}
so there are second order poles at $z=\pm z_{0}$. Plugging eqs \eqref{eq:mhv1}
and \eqref{eq:mhv2} into eq \eqref{eq:bcfw} then gives

\[
A_{4}^{MHV}=-\frac{1}{2\pi i}\frac{\delta^{8}(Q)}{\left\langle 12\right\rangle \left\langle 23\right\rangle \left\langle 34\right\rangle \left\langle 41\right\rangle }\left(1-z_{0}^{2}\right)^{2}\oint dz\frac{z^{2}}{z-1}\frac{1}{\left(z^{2}-z_{0}^{2}\right)^{2}}
\]
where the contour encircles the points $z=\pm z_{0}$. After evaluating
the integral, we obtain 
\[
A_{4}^{MHV}=\frac{\delta^{8}(Q)}{\left\langle 12\right\rangle \left\langle 23\right\rangle \left\langle 34\right\rangle \left\langle 41\right\rangle },
\]
which is the expected formula for the 4pt MHV amplitude.

\begin{figure}
\begin{center}
\includegraphics[scale=0.3]{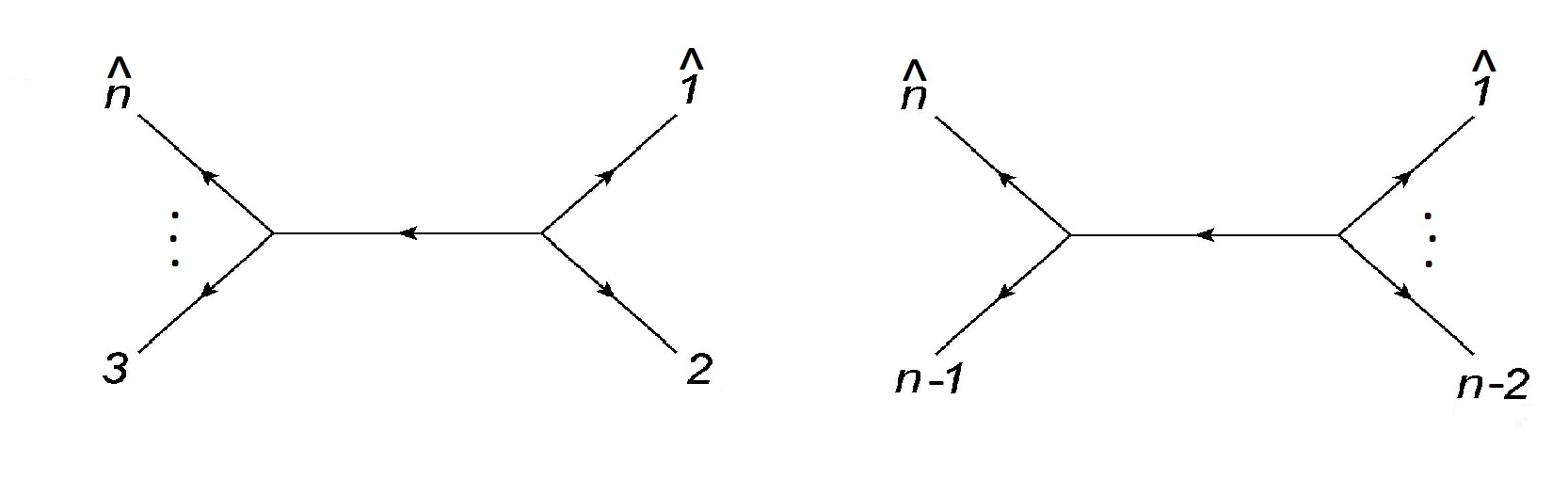}
\caption{BCFW diagrams for the n-pt MHV superamplitude. The $(n-1)$ pt amplitude in each factorization channel is an MHV superamplitude and the 3pt amplitude in each factorization channel is an anti-MHV superamplitude.}
\label{nptbcfw}
\end{center}
\end{figure}
Now let's generalize the previous calculation to obtain all $n$-pt MHV amplitudes.
If we deform legs 1 and $n$, then there are two diagrams we need to
consider, which are depicted in Figure \ref{nptbcfw}. For both diagrams, we find
that 
\begin{equation}
\int d^{4}\eta_{\hat{p}}\frac{A_{L}A_{R}}{\hat{p}^{2}}=\frac{\delta^{8}(Q)}{\left\langle \hat{1}2\right\rangle \left\langle 23\right\rangle \left\langle 34\right\rangle \ldots\left\langle n-2n-1\right\rangle \left\langle n-1\hat{n}\right\rangle \left\langle n1\right\rangle }.\label{eq:mhv3}
\end{equation}
This expression has two simple poles corresponding to the roots of $\left\langle \hat{1}2\right\rangle $
and two simple poles corresponding to the roots of $\left\langle n-1\hat{n}\right\rangle $.
The $n$-pt MHV amplitude is obtained by plugging eq \eqref{eq:mhv3}
into eq \eqref{eq:bcfw} and taking the contour to encircle each of these four poles.
The contribution from the roots of $\left\langle \hat{1}2\right\rangle $
correspond to the diagram on the left in Figure \ref{nptbcfw}, and the contribution
from the roots of $\left\langle n-1\hat{n}\right\rangle $ correspond
to the diagram on the right in Figure \ref{nptbcfw}. In particular, the contribution from the roots of $\left\langle \hat{1}2\right\rangle $ is
\begin{equation}
\frac{\delta^{8}(Q)}{\left\langle 12\right\rangle \left\langle 23\right\rangle \left\langle 34\right\rangle ...\left\langle n-2n-1\right\rangle \left\langle n1\right\rangle }\frac{\left\langle n2\right\rangle +i\left\langle 12\right\rangle }{\left\langle n2\right\rangle \left\langle n-1n\right\rangle +\left\langle n12\right\rangle \left\langle n-11\right\rangle }.\label{eq:npt1}
\end{equation}
and the contribution from the roots of $\left\langle n-1\hat{n}\right\rangle $ is
\begin{equation}
\frac{\delta^{8}(Q)}{\left\langle 23\right\rangle \left\langle 34\right\rangle ...\left\langle n-2n-1\right\rangle \left\langle n-1n\right\rangle \left\langle n1\right\rangle }\frac{\left\langle n-11\right\rangle -i\left\langle n-1n\right\rangle }{\left\langle n2\right\rangle \left\langle n-1n\right\rangle +\left\langle n12\right\rangle \left\langle n-11\right\rangle }.\label{eq:npt2}
\end{equation}
Adding eqs \eqref{eq:npt1} and \eqref{eq:npt2} gives
\[
A_{n}^{MHV}=\frac{\delta^{8}(Q)}{\left\langle 12\right\rangle \left\langle 23\right\rangle ...\left\langle n1\right\rangle }
\]
which is indeed the $n$-pt MHV superamplitude.   

\section{R-symmetry of $\mathcal{N}=8$ sYM Amplitudes \label{rsymm}}

When dimensionally reducing 4d $\mathcal{N}=4$ sYM to 3d, one component
of the 4d gauge field becomes a scalar in three-dimensions. Since
$\mathcal{N}=4$ sYM has has six scalars, it follows that $\mathcal{N}=8$
sYM has seven scalars. Furthermore, the Lagrangian of $\mathcal{N}=8$
sYM has $SO(7)$ R-symmetry which rotates the seven scalars. As explained
in section \ref{4d23d}, we must break the $SO(7)$ R-symmetry to $SU(4)$ in order to define helicity structure in $\mathcal{N}=8$ sYM. In this appendix, we will explain how $SO(7)$ R-symmetry is realized
in the amplitudes of $\mathcal{N}=8$ sYM. 

In section \ref{4d23d}, we defined the following $SU(4)$ R-symmetry generators:

\[
r_{\,\,\,\, J}^{I}=\sum_{i=1}^{n}\left(-\eta_{i}^{I}\frac{\partial}{\partial\eta_{i}^{J}}+\frac{1}{4}\delta_{J}^{I}\eta_{i}^{K}\frac{\partial}{\partial\eta_{i}^{K}}\right).\]
where $I,J=1,2,3,4$ are $SU(4)$ indices and $i$ labels the external
legs of an $n$-pt amplitude. Let us now consider the following generators:

\begin{equation}
r^{IJ}=\sum_{i=1}^{n}\eta_{i}^{I}\eta_{i}^{J},\,\,\, r_{IJ}=\sum_{i=1}^{n}\frac{\partial}{\partial\eta_{i}^{I}}\frac{\partial}{\partial\eta_{i}^{J}},\,\,\, r=\sum_{i=1}^{n}\eta_{i}^{I}\frac{\partial}{\partial\eta_{i}^{I}}{\normalcolor .}\label{eq:rs}\end{equation}
When combined with $r_{\,\,\,\, J}^{I}$, these generate the group
$SO(8)$. It is not difficult to see that $r_{IJ}$ annihilates the
MHV amplitudes. In particular,\begin{equation}
r_{IJ}A_{n}^{MHV}=\frac{\delta^{3}(P)}{\left\langle 12\right\rangle \left\langle 23\right\rangle ...\left\langle n1\right\rangle }\sum_{i=1}^{n}\frac{\partial}{\partial\eta_{i}^{I}}\frac{\partial}{\partial\eta_{i}^{J}}\delta^{8}(Q)=\frac{\delta^{3}(P)}{\left\langle 12\right\rangle \left\langle 23\right\rangle ...\left\langle n1\right\rangle }\sum_{i=1}^{n}\lambda_{i}^{\alpha}\lambda_{i}^{\beta}\frac{\partial}{\partial Q^{I\alpha}}\frac{\partial}{\partial Q^{J\beta}}\delta^{8}(Q)=0\label{eq:r_ij}\end{equation}
where we used the chain rule $\frac{\partial}{\partial\eta_{i}^{I}}=\frac{\partial Q^{K\alpha}}{\partial\eta_{i}^{I}}\frac{\partial}{\partial Q^{K\alpha}}=\lambda_{i}^{\alpha}\frac{\partial}{\partial Q^{I\alpha}}$, 
and noted that $\sum_{i=1}^{n}\lambda_{i}^{\alpha}\lambda_{i}^{\beta}$
vanishes on the support of $\delta^{3}(P)$. Furthermore, if we note
that $N^{n-4}$MHV amplitudes can be obtained by Fourier transforming anti-MHV amplitudes with respect to the $\bar{\eta}$ variables,
it follows that $r^{IJ}$ annihilates the $N^{n-4}$MHV amplitudes.
In particular, \[
A_{n}^{N^{n-4}MHV}=\Pi_{i=1}^{n}\Pi_{I=1}^{4}\int d\bar{\eta}_{Ii}e^{\eta_{i}^{I}\bar{\eta}_{Ii}}\frac{\delta^{3}(P)\delta^{8}\left(\bar{Q}\right)}{\left\langle 12\right\rangle \left\langle 23\right\rangle ...\left\langle n1\right\rangle }\]
where $\bar{Q}_{I\alpha}=\left(Q^{I\alpha}\right)^{*}=\sum_{i=1}^{n}\bar{\eta}_{Ii}\lambda_{i}^{\alpha}$, 
and the indices $i$ and $I$ appearing in the integrand are not summed over.
Taking the complex conjugate of eq \eqref{eq:r_ij} and Fourier transforming
then gives\[
0=\Pi_{i=1}^{n}\Pi_{K=1}^{4}\int d\bar{\eta}_{Ki}e^{\eta_{i}^{K}\bar{\eta}_{Ki}}\sum_{j=1}^{n}\frac{\partial}{\partial\bar{\eta}_{Ij}}\frac{\partial}{\partial\bar{\eta}_{Jj}}\frac{\delta^{3}(P)\delta^{8}\left(\bar{Q}\right)}{\left\langle 12\right\rangle \left\langle 23\right\rangle ...\left\langle n1\right\rangle }=r^{IJ}A_{n}^{N^{n-4}MHV}\]
where we performed integration by parts to get the last equality. 

From the discussion above, we find that $r_{IJ}$ is a symmetry of
the MHV amplitudes and $r^{IJ}$ is a symmetry of the $N^{n-4}$MHV
amplitudes. It then follows that $r_{IJ}$ and $r^{IJ}$ are both symmetries
of the 4pt amplitude. Furthermore, if we note the generator $r$ in
eq \eqref{eq:rs} simply counts the fermionic degree of an amplitude,
we see that $r-8$ is also symmetry of the 4pt amplitude. This implies
that the 4pt amplitude has $SO(8)$ R-symmetry, which confirms the
results of \cite{Agarwal:2011tz}. One may ask
whether higher-point amplitudes also have this symmetry. We will now
show that $SO(8)$ R-symmetry does not extend to higher point amplitudes
since $r_{IJ}$ and $r^{IJ}$ are not symmetries of generic MHV or
anti-MHV amplitudes.
For example, let's consider the action of $r^{IJ}$ on an MHV amplitude.
Taking $I=1$ and $J=2$ gives: \[
r^{12}A_{n}^{MHV}=\frac{\delta^{3}(P)Q^{3\alpha}Q_{\alpha}^{3}Q^{4\beta}Q_{\beta}^{4}}{\left\langle 12\right\rangle \left\langle 23\right\rangle ...\left\langle n1\right\rangle }\sum_{i=1}^{n}\eta_{i}^{1}\eta_{i}^{2}Q^{1\gamma}Q_{\gamma}^{1}Q^{2\delta}Q_{\delta}^{2}\]
where we noted that $\delta^{8}(Q)=\Pi_{K=1}^{4}Q^{K\alpha}Q_{\alpha}^{K}$.
Expanding $\sum_{i=1}^{n}\eta_{i}^{1}\eta_{i}^{2}Q^{1\alpha\gamma}Q_{\gamma}^{1}Q^{2\delta}Q_{\delta}^{2}$
will then give a sum of terms, each of which has fermionic degree six. For
example, there is a term proportional to $\eta_{1}^{1}\eta_{2}^{1}\eta_{3}^{1}\eta_{1}^{2}\eta_{2}^{2}\eta_{3}^{2}$
and the coefficient of this term is $\left\langle 12\right\rangle ^{2}+\left\langle 13\right\rangle ^{2}+\left\langle 23\right\rangle ^{2}$.
When $n=4$, the the sum of these three kinematic invariants vanishes,
but this is not generally true for $n>4$. As another example, let's
consider the term proportional to $\eta_{1}^{1}\eta_{2}^{1}\eta_{3}^{1}\eta_{1}^{2}\eta_{2}^{2}\eta_{4}^{2}$.
The coefficient of this term is given by $\left\langle 23\right\rangle \left\langle 24\right\rangle +\left\langle 13\right\rangle \left\langle 14\right\rangle $.
When $n=4$, momentum conservation implies that \[
\frac{\left\langle 21\right\rangle }{\left\langle 34\right\rangle }=\frac{\left\langle 23\right\rangle }{\left\langle 14\right\rangle }=\frac{\left\langle 13\right\rangle }{\left\langle 42\right\rangle }=\pm1\]
so $\left\langle 23\right\rangle \left\langle 24\right\rangle +\left\langle 13\right\rangle \left\langle 14\right\rangle =0$,
but this not generally true when $n>4$. Hence, we find that \[
r^{IJ}A_{n>4}^{MHV}\neq0{\normalcolor .}\]
Taking the complex conjugate of the above equation and Fourier transforming
with respect to the $\bar{\eta}$ variables then implies that\[
r_{IJ}A_{n>4}^{N^{n-4}MHV}\neq0{\normalcolor .}\]

In summary, we have found that when $n>4$, the amplitudes of $\mathcal{N}=8$
sYM do not have $SO(8)$ symmetry. On the other hand, since the action
of $\mathcal{N}=8$ sYM has manifest $SO(7)$ R-symmetry, the amplitudes
must also have $SO(7)$ R-symmetry.
Noting that $r^{IJ}$ generally increases the fermionic degree of
an amplitude by two, and $r_{IJ}$ generally decreases the fermionic
degree of an amplitude by two, it follows that $\tilde{r}^{IJ}=r^{IJ}+\frac{1}{2}\epsilon^{IJKL}r_{KL}$
should be a symmetry of the amplitudes. For each $n$, we can construct
the following amplitude: \[
A_{n}=\sum_{k=0}^{n-4}\alpha_{k}A_{n}^{N^{k}MHV}\]
where $\alpha_{k}$ are coefficients chosen such that $\tilde{r}^{IJ}A_{n}=0$.
In particular, $\tilde{r}^{IJ}A_{n}$ should vanish as a telescoping
sum. It would be interesting to compute the $\alpha_k$ explicitly for $n>4$. When combined with $r_{\,\,\,\, J}^{I}$, $\tilde{r}^{IJ}$ generates the group $SO(7)$. Hence, we conjecture that for each $n$, there
is a single amplitude which has $SO(7)$ R-symmetry (but not $SO(8)$
R-symmetry when $n>4$). In the IR, the R-symmetry of these amplitudes
should be enhanced to $SO(8)$, and they should match onto
the amplitudes of the BLG theory, which has manifest $SO(8)$
R-symmetry.

\section{Tree-Level Dual Conformal Symmetry  \label{dualtree}}

In this appendix, we will prove that the tree-level amplitudes of $\mathcal{N}=8$ sYM have dual conformal symmetry when they are stripped of a supermomentum delta function. Decomposing the amplitudes according to eq \eqref{eq:decomp}, the BCFW recursion relation in eq \eqref{eq:bcfw} can be written as follows:

\begin{equation}
f_{n}=-\frac{1}{2\pi i}\sum_{f,j}\oint_{z_{f,j}}\int d^{4}\eta\delta^{8}\left(Q_{R}\right)\frac{f_{L}(z,\eta)f_{R}(z,i\eta)}{\hat{p}_{f}(z)^{2}}\frac{1}{z-1}
\label{eq:fbc}
\end{equation}  
where the sum is over factorization channels and the poles in each factorization channel. For each channel, $Q_R$ is the sum of the supermomenta of the amplitude on the right-hand side. Assuming that $f_{k}$
inverts covariantly for $k<n$, we will show that $f_{n}$ must invert
covariantly as well by performing a dual inversion of eq \eqref{eq:fbc}.
Let's choose legs $1$ and $n$ to be deformed according to eq \eqref{eq:deform}.
Note that this is equivalent to deforming $\left(x_{1},\theta_{1}\right)$
in the dual space since from eq \eqref{eq:hyper}
we have

\[
\hat{p}_{1}=\hat{x}_{1}-x_{2},\,\,\,\hat{p}_{n}=x_{n}-\hat{x}_{1},\,\,\,\hat{q}_{1}=\hat{\theta}_{1}-\theta_{n},\,\,\,\hat{q}_{n}=\theta_{n}-\hat{\theta}_{1}{\normalcolor .}\]
Since the amplitudes are invariant under translations in the dual
space, let us choose the origin of the dual space such that

\begin{equation}
x_{2}=-x_{n}{\normalcolor .}\label{eq:choice}\end{equation}
Then $p_{1}=x_{1}-x_{2}=x_{1}+x_{n}$ and $p_{n}=x_{n}-x_{1}$,
which implies that $x_{1}=\frac{1}{2}\left(p_{1}-p_{n}\right)$. It
then follows that $x_{1}^{2}$ is invariant under the deformation
in eq \eqref{eq:deform}, since \[
\hat{x}_{1}^{2}=-\frac{1}{2}\hat{p}_{1}\cdot\hat{p}_{n}=-\frac{1}{4}\left(\hat{p}_{1}+\hat{p}_{n}\right)^{2}=-\frac{1}{4}\left(p_{1}+p_{n}\right)^{2}=-\frac{1}{2}p_{1}\cdot p_{n}=x_{1}^{2}{\normalcolor .}\]
This is why eq \eqref{eq:choice} is a convenient choice. 

\begin{figure}
\begin{center}
\includegraphics[scale=0.2]{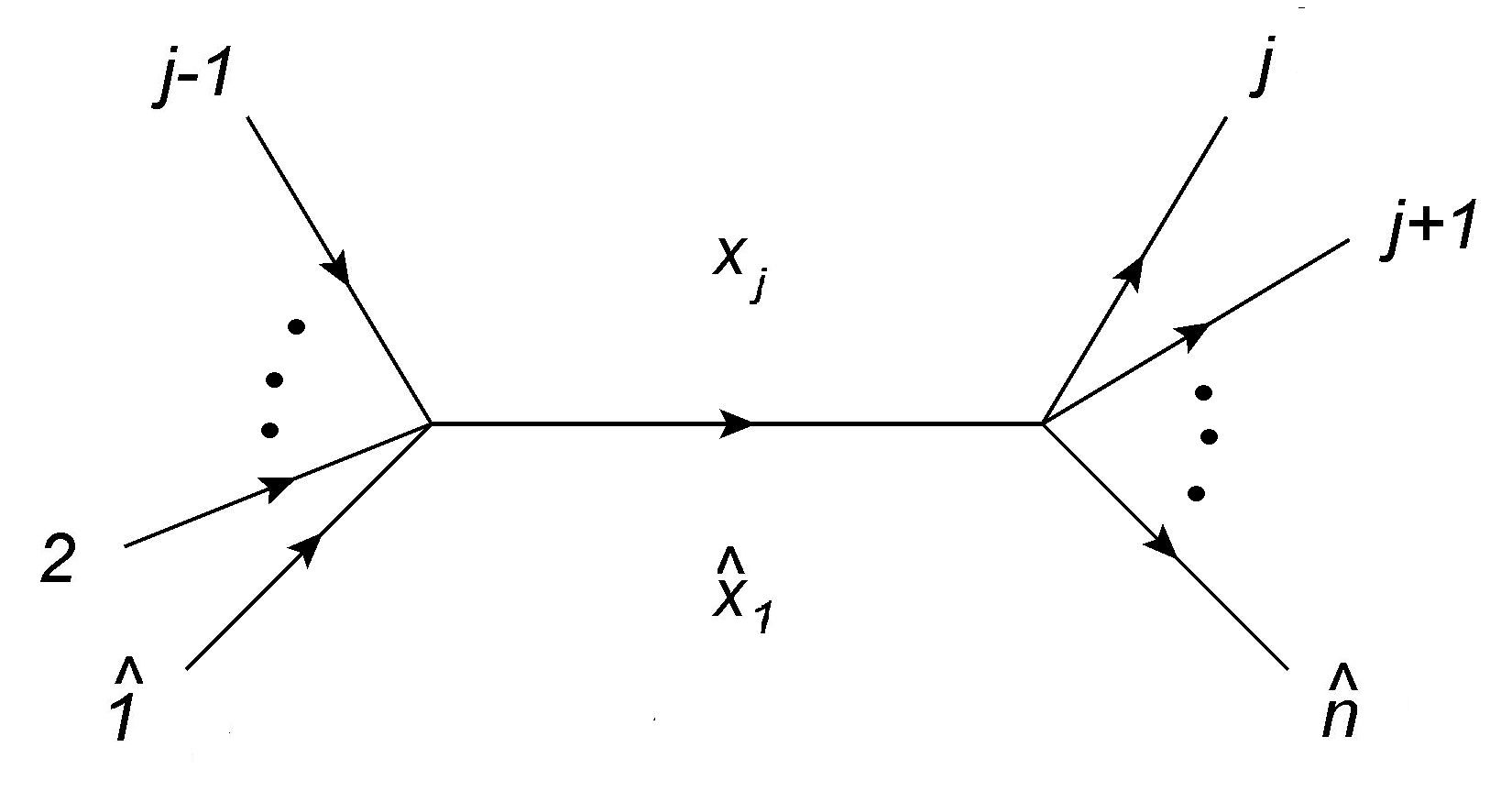}
\caption{ A factorization channel in the BCFW recursion relation where legs $1$ and $n$ are shifted. Dual space coordinates are indicated.}
\label{dualtreediag}
\end{center}
\end{figure}

For simplicity, let's focus on one channel in eq \eqref{eq:fbc}. In
particular, let $f_{L}$ correspond to a $j$-point amplitude and
$f_{R}$ correspond to a $k$-point amplitude where $k=n-j+2$ as depicted in Figure \ref{dualtreediag}. Assuming that the subamplitudes invert covariantly,
we have

\begin{equation}
I\left[f_{L}\left(\hat{x}_{1},...,x_{j}\right)\right]=x_{1}^{2}...x_{j}^{2}f_{L}\left(\hat{x}_{1},...,x_{j}\right),\,\,\, I\left[f_{R}\left(x_{j},...,x_{n},\hat{x}_{1}\right)\right]=x_{j}^{2}...x_{n}^{2}x_{1}^{2}f_{R}\left(x_{j},...,x_{n},\hat{x}_{1}\right){\normalcolor .}\label{eq:dual1}\end{equation}

Now let's look at how the $\int d^{4}\eta\delta^{8}\left(Q_{R}\right)$
transforms under dual inversions. Noting that $Q_{R}^{I\alpha}=\theta_{j}^{I\alpha}-\hat{\theta}_{1}^{I\alpha}-\eta^{I}\hat{\lambda}_{f}^{\alpha}$,
where $\hat{\lambda}_{f}^{\alpha}\hat{\lambda}_{f}^{\beta}=\hat{p}_{f}^{\alpha\beta}(z)$
near its roots, we find that\[
\int d^{4}\eta\delta^{8}\left(Q_{R}\right)=\frac{1}{4!}\epsilon_{IJKL}\left(\theta_{j}-\hat{\theta}_{1}\right)^{I\alpha}\hat{\lambda}_{f\alpha}\left(\theta_{j}-\hat{\theta}_{1}\right)^{J\beta}\hat{\lambda}_{f\beta}\left(\theta_{j}-\hat{\theta}_{1}\right)^{K\gamma}\hat{\lambda}_{f\gamma}\left(\theta_{j}-\hat{\theta}_{1}\right)^{L\delta}\hat{\lambda}_{f\delta}{\normalcolor .}\]
Furthermore using eq \eqref{eq:inverse} and the fact that $I\left[\lambda_{i}\right]=\frac{x_{i}\lambda_{i}}{\sqrt{x_{i+1}^{2}x_{i}^{2}}}$ \cite{Huang:2010qy},
we obtain\[
I\left[\left(\theta_{j}-\hat{\theta}_{1}\right)^{I\alpha}\hat{\lambda}_{f\alpha}\right]=\frac{\left(\theta_{j}-\hat{\theta}_{1}\right)^{I\alpha}\hat{\lambda}_{f\alpha}}{\sqrt{x_{1}^{2}x_{j}^{2}}}\]
from which it follows that
\begin{equation}
I\left[\int d^{4}\eta\delta^{8}\left(Q_{R}\right)\right]=\frac{1}{\left(x_{1}^{2}x_{j}^{2}\right)}\int d^{4}\eta\delta^{8}\left(Q_{R}\right){\normalcolor .}\label{eq:dual3}
\end{equation}

Finally, note that 
\begin{equation}
I\left[\hat{p}_{f}(z)^{2}\right]=\frac{\hat{p}_{f}(z)^{2}}{x_{1}^{2}x_j^{2}},
\label{pinv}
\end{equation}
Combining eqs \eqref{eq:dual1}, \eqref{eq:dual3}, and \eqref{pinv},
we see that performing a dual inversion of eq \eqref{eq:fbc} indeed
implies eq \eqref{eq:state}. 

\section{Dual Conformal Covariance of Loop Integrands \label{dualloop}}

\begin{figure}
\begin{center}
\includegraphics[scale=0.9]{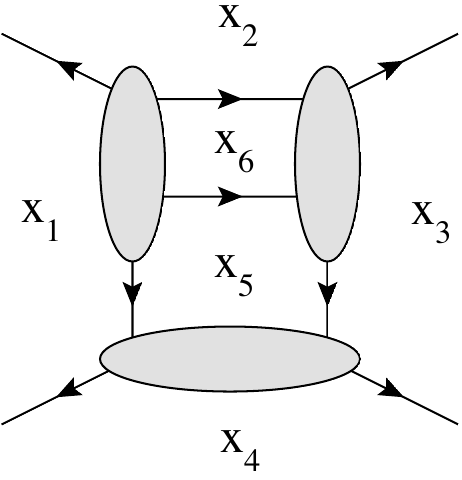}
\caption{A sample unitarity cut of four-point two-loop amplitude. }
\label{dualloopdiag}
\end{center}
\end{figure}

In this appendix, we will use the results of appendix \ref{dualtree} and unitarity to show that the cut constructable loop integrands of $\mathcal{N}=8$ sYM transform covariantly under dual inversions. The argument is very similar to the one used to prove loop-level dual conformal symmetry in the ABJM theory \cite{Gang:2010gy}. For loop amplitudes, the labeling of the dual space cannot always be done such that adjacent regions are labeled successively. For example, $x_6$ and $x_2$ in Fig \ref{dualloopdiag} are non-successive yet they are adjacent regions. Thus we utilize the following more general notation:
\begin{equation}
x^{\alpha\beta}_{i}-x^{\alpha\beta}_{j}=p^{\alpha\beta}_{\{ij\}},\;\;\theta^{\alpha I}_{i}-\theta^{\alpha I}_{j}=q^{\alpha I}_{\{ij\}}\,.
\end{equation}

The cut equation is given by
\begin{eqnarray}
 \mathcal{A}_n^{L}\Bigr|_{\hbox{\footnotesize{cut}}} &=& \int
    \prod_{\{ij\}} d^4\eta_{\{ij\}}  \prod_{\alpha}
    \mathcal{A}^{{\rm tree}}_{\alpha}
\nn\\
  &=& \delta^3(P) \int
   \prod_{\{ij\}} d^4\eta_{\{ij\}} \times
    \prod_{\alpha} \delta^8\left(Q_{\alpha}\right) f_{\alpha}
\end{eqnarray}
where $\{ij\}$ runs over cut lines and $\alpha$ runs over the tree diagrams in the cut. In obtaining the second line, we used the decomposition for tree amplitudes described in the previous subsection, notably $ \mathcal{A}^{{\rm tree}}_{\alpha} = \delta^3(P_\alpha)\delta^8\left(Q_{\alpha}\right) f_{\alpha}$  and noted that the product of the momentum delta functions from each tree amplitude gives rise to an overall momentom delta function.  

Now let's convert the $\eta$ integrals to integrals over $\theta$'s. In order to do so, we need to use the following identity:
\begin{equation}
 \prod_{\alpha} \delta^8\left(Q_{\alpha}\right)=\int \left[\prod_k d^8\theta_k\right]\left[ \prod_{\{rs\}}\delta^8\left(\theta_r-\theta_s-\lambda_{\{rs\}}\eta_{\{rs\}}\right)\right]
 \label{transformer}
\end{equation}
where $k$ runs over all regions in the dual space except one external region and $\{rs\}$ runs over all cut and external lines. Eq \eqref{transformer} can be understood using the following argument. The total number of regions in the dual space is $F=n+L$, where $n$ is the number of external legs and $L$ is the loop level. Since the amplitudes have an overall shift symmetry when written in dual space coordinates, we can set $\theta$ in one of the external regions of the dual space to zero, so the integration measure $d^8\theta_k$ only includes $F-1$ of the regions in dual space.  Hence there are $8(F-1)$ integrals on the right hand side of equation \eqref{transformer}.  Denoting the total number of cut and external lines as $P$, there are therefore $8(P-F+1)$ delta functions remaining on the right hand side after performing these integrals. On the other hand, for planar diagrams $P-F+1=V$, where $V$ is the number of tree amplitudes in the cut.  In Fig.\ref{dualloopdiag}, for example, $P=8$, $F=6$, and $V=3$. Since the index $\alpha$ on the left hand side of equation \eqref{transformer} runs over all tree diagrams in the cut, the number of supermomentum delta functions on each side of the equation is indeed the same. 

Using eq.(\eqref{transformer}), one can rewrite the cut equation as:
\begin{eqnarray}
   \nn\\
  \mathcal{A}_n^{L}\Bigr|_{\hbox{\footnotesize{cut}}} &=&  \delta^3(P)\int
    \prod_{\{ij\}} d^4\eta_{\{ij\}} \times
    \prod_k d^8\theta_k \times
    \prod_\alpha f_\alpha \nn\\
  &&\times\prod_{\{rs\}}\delta^8\left(\theta_r-\theta_s-\lambda_{\{rs\}}\eta_{\{rs\}}\right)
     \,
\label{loopamplitude}
\end{eqnarray}
where once again $\{ij\}$ runs over cut lines, $k$ runs over $n+L-1$ regions in the dual space, $\alpha$ runs over the tree diagrams in the cut, and $\{rs\}$ runs over all cut and external lines. After using the delta functions to eliminate the $\eta_{\{ij\}}$-dependence from each $f_\alpha$, the $\eta_{\{ij\}}$-dependence comes solely from the delta functions. The integral over $\eta_{\{ij\}}$ then simplifies to
\begin{eqnarray}
  \int d^4\eta_{\{ij\}} \delta^8\left(\theta_i-\theta_j-\lambda_{\{ij\}}\eta_{\{ij\}}\right)=\delta^4\left(\theta_{ij}\lambda_{\{ij\}}\right) ,\,
\end{eqnarray}
where $\theta_{ij}\lambda_{\{ij\}}=(\theta_i-\theta_j)^{\alpha}\lambda_{\alpha\{ij\}}$, and we have suppressed the SU(4) R-index. Hence, performing the $\eta_{ij}$ integrals  in eq \eqref{loopamplitude} gives
\begin{eqnarray}
  \mathcal{A}_n^{L}\Bigr|_{\hbox{\footnotesize{cut}}} &=& \delta^3(P) \int
    \prod_k d^8\theta_k\times
    \prod_\alpha f_\alpha \times
    \prod_{\{ij\}} \delta^4\left(\theta_{ij}\lambda_{\{ij\}}\right)\nn\\
  &&\times\prod_{\{rs\}}\delta^8\left(\theta_r-\theta_s-\lambda_{\{rs\}}\eta_{\{rs\}}\right) \,,
\end{eqnarray}
where $\{rs\}$ now only runs over the external lines, i.e. there are $n$ of them. Furthermore, one can pull out an overall supermomentum delta function, leaving $(n-1)$ delta functions which trivialize the integrals over the $\theta$ coordinates of $n-1$ external regions of the dual space. After performing these $n-1$ integrals, we are finally left with
\begin{eqnarray}
  \mathcal{A}_n^{L}\Bigr|_{\hbox{\footnotesize{cut}}}=\delta^3(P)\delta^8(Q) \int \left(\prod_k d^8 \theta_k \right)
  \prod_{\{ij\}} \delta^4\left(\theta_{ij}\lambda_{\{ij\}}\right)
  \prod_\alpha f_\alpha \,,
\label{cutintegrand}
\end{eqnarray}
where now $k$ now runs over the loop regions in the dual space, which are regions 5 and 6 in the example shown in Fig.\ref{dualloop}.

Let us consider the inversion weight of each term in eq.(\eqref{cutintegrand}):
\begin{itemize}
  \item For each loop region $k$, the $\theta_k$ measure contributes a factor $(x_k^2)^4$.
  \item Each cut leg $\{ij\}$ contributes $(x_i^2x_j^2)^{-2}$, which comes from $\delta^4\left(\theta_{ij}\lambda_{\{ij\}}\right)$.
  \item Each tree-level sub-amplitude contributes $\prod_i x_i^2$, where $i$ runs over all regions adjacent to the tree.
\end{itemize}
Furthermore, after restoring the cut propagators, which invert as
\begin{equation}
I\left[\frac{1}{p^2_{\{ij\}}}\right]=I\left[\frac{1}{x^2_{ij}}\right]=\frac{x_i^2x_j^2}{x^2_{ij}}\,,
\end{equation}
one can show that the integrand of a loop diagram with $n$ external legs and $L$ loops must invert as follows under a dual inversion:
\begin{equation}
I\left[\mathcal{I}_{n}^{L}\right]=\Pi_{i=1}^{n}x_{i}^{2}\Pi_{j=1}^{L}\left(x_{j}^{2}\right)^{4}\mathcal{I}_{n}^{L},
\label{integrandinversion}
\end{equation}
where $i$ runs over the external regions and $j$ runs over loop regions in the dual space.

As a concrete example, consider the diagram in Fig.\ref{dualloopdiag}. One has
\begin{eqnarray}
\nonumber I\left[ \mathcal{A}^{\rm Fig.\ref{dualloopdiag}}\Bigr|_{\hbox{\footnotesize{cut}}}\right]&=&\frac{(x_5^2x_6^2)^4 (x_1^2x_2^2x^2_6x^2_5)(x_1^2x_3^2x^2_4x^2_5)(x_3^2x_2^2x^2_6x^2_5)}{(x_2^2x_6^2x_6^2x_5^2x_5^2x_1^2x_5^2x_3^2)^2}\mathcal{A}^{\rm Fig.\ref{dualloopdiag}}\Bigr|_{\hbox{\footnotesize{cut}}}\\
&=& x^2_4 x^2_5 (x^2_6)^2\mathcal{A}^{\rm Fig.\ref{dualloopdiag}}\Bigr|_{\hbox{\footnotesize{cut}}}.
\end{eqnarray}
Furthermore, there are four propagators in Fig.\ref{dualloopdiag}, which invert as follows:
\begin{equation}
I\left[\frac{1}{x^2_{15}}\right]=\frac{x_1^2x^2_5}{x^2_{15}},\;\;I\left[\frac{1}{x^2_{53}}\right]=\frac{x_3^2x^2_5}{x^2_{53}},\;\;I\left[\frac{1}{x^2_{56}}\right]=\frac{x_6^2x^2_5}{x^2_{56}},\;\;I\left[\frac{1}{x^2_{62}}\right]=\frac{x_2^2x^2_6}{x^2_{62}}.
\end{equation}
Thus, when $\mathcal{A}^{\rm Fig.\ref{dualloopdiag}}\Bigr|_{\hbox{\footnotesize{cut}}}$ is combined with the cut propagators, the resulting object has the following inversion weight:
\begin{equation}
 x_1^2x_2^2x_3^2x^2_4 (x^2_5)^4(x^2_6)^4\,,
\end{equation}
which matches the result in eq.(\eqref{integrandinversion})

\section{Proof of eq \eqref{eq:lambdaprop}  \label{momtwistproof}}

In this appendix, we will relate the spinors $\tilde{\lambda}$ of 3d momentum twistors (which appear in eq \eqref{eq:3dx}) to the spinors $\lambda$ of the on-shell superspace (which appear in eq \eqref{eq:hyper}). Using eq \eqref{eq:3dx} and the Schouten identity, it is easy to show that

\begin{equation}
\left(x_{i}-x_{i+1}\right)^{\alpha\beta}=\tilde{\lambda}_{i}^{\alpha}\frac{\left\langle \tilde{i}-1\tilde{i}+1\right\rangle \tilde{\mu}_{i}^{\beta}+\left\langle \tilde{i}+1\tilde{i}\right\rangle \tilde{\mu}_{i-1}^{\beta}-\left\langle \tilde{i}-1\tilde{i}\right\rangle \tilde{\mu}_{i+1}^{\beta}}{\left\langle \tilde{i}1\tilde{i}-1\right\rangle \left\langle \tilde{i}+1\tilde{i}\right\rangle }{\normalcolor .}\label{eq:3dx-1}\end{equation}
Consider the first term in the numerator, $\left\langle \tilde{i}-1\tilde{i}+1\right\rangle \tilde{\mu}_{i}^{\beta}$. It can be written as follows:
\begin{equation}
\left\langle \tilde{i}-1\tilde{i}+1\right\rangle \tilde{\mu}_{i}^{\beta}=\frac{1}{2}\left(\tilde{\mu}_{i}^{\beta}\tilde{\lambda}_{i-1}^{\gamma}\tilde{\lambda}_{i+1\gamma}-\tilde{\mu}_{i}^{\beta}\tilde{\lambda}_{i+1}^{\gamma}\tilde{\lambda}_{i-1\gamma}\right){\normalcolor .}\label{eq:appedix1}\end{equation}
The first term on RHS of eq \eqref{eq:appedix1} can be
written as 
\[
\mbox{\ensuremath{\tilde{\mu}_{i}^{\beta}\tilde{\lambda}_{i-1}^{\gamma}}}\tilde{\lambda}_{i+1\gamma}=-\frac{1}{2}\epsilon^{\beta\gamma}\tilde{\mu}_{i}\cdot\tilde{\lambda}_{i-1}\tilde{\lambda}_{i+1}^{\beta}+\frac{1}{2}\tilde{\mu}_{i}^{\gamma}\tilde{\lambda}_{i-1}^{\beta}\tilde{\lambda}_{i+1\gamma}
\]
and the second term on RHS can be written as 
\[
\tilde{\mu}_{i}^{\beta}\tilde{\lambda}_{i+1}^{\gamma}\tilde{\lambda}_{i-1\gamma}=-\frac{1}{2}\tilde{\mu}_{i}\cdot\tilde{\lambda}_{i+1}\tilde{\lambda}_{i-1}^{\beta}+\frac{1}{2}\tilde{\mu}_{i}^{\gamma}\tilde{\lambda}_{i+1}^{\beta}\tilde{\lambda}_{i-1\gamma}.
\]
Plugging these expressions into the RHS of eq \eqref{eq:appedix1} then
gives \[
\left\langle \tilde{i}-1\tilde{i}+1\right\rangle \tilde{\mu}_{i}^{\beta}=\frac{1}{2}\tilde{\mu}_{i}\cdot\tilde{\lambda}_{i+1}\tilde{\lambda}_{i-1}^{\beta}-\frac{1}{2}\tilde{\mu}_{i}\cdot\tilde{\lambda}_{i-1}\tilde{\lambda}_{i+1}^{\beta}+\frac{1}{2}\left\langle \tilde{i}-1\tilde{i}+1\right\rangle \tilde{\mu}_{i}^{\beta}{\normalcolor ,}\]
which implies that\[
\left\langle \tilde{i}-1\tilde{i}+1\right\rangle \tilde{\mu}_{i}^{\beta}=\tilde{\mu}_{i}\cdot\tilde{\lambda}_{i+1}\tilde{\lambda}_{i-1}^{\beta}-\tilde{\mu}_{i}\cdot\tilde{\lambda}_{i-1}\tilde{\lambda}_{i+1}^{\beta}.\]

From the constraint in eq \eqref{eq:3dconstraint}, we have $\tilde{\mu}_{i}\cdot\tilde{\lambda}_{i+1}=\tilde{\mu}_{i+1}\cdot\tilde{\lambda}_{i}$ and 
$\tilde{\mu}_{i}\cdot\tilde{\lambda}_{i-1}=\tilde{\mu}_{i-1}\cdot\tilde{\lambda}_{i}$,
which implies that

\[
\left\langle \tilde{i}-1\tilde{i}+1\right\rangle \tilde{\mu}_{i}^{\beta}=\tilde{\lambda}_{i\gamma}\left(\tilde{\lambda}_{i-1}^{\beta}\tilde{\mu}_{i+1}^{\gamma}-\tilde{\lambda}_{i+1}^{\beta}\tilde{\mu}_{i-1}^{\gamma}\right){\normalcolor .}\]
Plugging this into the numerator of eq \eqref{eq:3dx-1} then gives

\[
\left\langle \tilde{i}-1\tilde{i}+1\right\rangle \tilde{\mu}_{i}^{\beta}+\left\langle \tilde{i}+1\tilde{i}\right\rangle \tilde{\mu}_{i-1}^{\beta}-\left\langle \tilde{i}-1\tilde{i}\right\rangle \tilde{\mu}_{i+1}^{\beta}=\tilde{\lambda}_{i}^{\beta}\left[\tilde{i}-1\tilde{i}+1\right]{\normalcolor .}\]
Hence, eq \eqref{eq:3dx-1} reduces to

\[
\left(x_{i}-x_{i+1}\right)^{\alpha\beta}=\frac{\left[\tilde{i}-1\tilde{i}+1\right]}{\left\langle \tilde{i}1\tilde{i}-1\right\rangle \left\langle \tilde{i}+1\tilde{i}\right\rangle }\tilde{\lambda}_{i}^{\alpha}\tilde{\lambda}_{i}^{\beta}{\normalcolor .}\]
Comparing this to eq \eqref{eq:hyper}, we see that \[
\lambda_{i}^{\alpha}=\sqrt{\frac{\left[\tilde{i}-1\tilde{i}+1\right]}{\left\langle \tilde{i}\tilde{i}-1\right\rangle \left\langle \tilde{i}+1\tilde{i}\right\rangle }}\tilde{\lambda}_{i}^{\alpha}{\normalcolor .}\]

\section{1-Loop Box Integrals  \label{loops}}

In this appendix we compute the scalar box integrals  $I_{4}^{4m},I_{4}^{3m}, I_{4}^{2mh}, I_{4}^{2me}, I_{4}^{1m}$, and $I_{4}^{0m}$ in 3d. These box integrals are defined in section \ref{3dloops}. In particular, they correspond to the box diagram in Fig \ref{boxfig} with massless propagators and various external legs taken to be massless.  

In \cite{Duplancic:2000sk}, $I_{4}^{3m}, I_{4}^{2mh}, I_{4}^{2me}, I_{4}^{1m}$, and $I_{4}^{0m}$ were computed in $4+2\epsilon_{IR}$ dimensions using the Feynman parameter approach. In this appendix, we will adapt the calculation of reference \cite{Duplancic:2000sk} to three dimensions by taking formulas in \cite{Duplancic:2000sk} which are valid for any value of $\epsilon_{IR}$, replacing $\epsilon_{IR}$ with $-1/2+\tilde{\epsilon}$, and expanding about $\tilde{\epsilon}=0$.  

\subsection{4m box}

In three dimensions, $I_{4}^{4m}$
can be reduced to the following integral over Feynman parameters: 

\[
I_{4}^{4m}=\frac{i}{(4\pi)^{3/2}}\Gamma\left(\frac{5}{2}\right)\int_{0}^{1}dx_{1}\int_{0}^{1-x_{1}}dx_{2}\int_{0}^{1-x_{1}-x_{2}}dx_{3}\left[\left(1-x_{1}-x_{2}-x_{3}\right)\left(x_{1}m_{4}^{2}+x_{2}t+x_{3}m_{3}^{2}\right)\right.\]
\[
\left.+x_{1}x_{2}m_{1}^{2}+x_{1}x_{3}s+x_{2}x_{3}m_{2}^{2}+i\epsilon\right]^{-5/2}\]
where we have set the renormalization scale $\mu=1$. If we first
make the substitution $x_{3}\rightarrow\left(1-x_{1}-x_{2}\right)x_{3}$
and then $x_{2}\rightarrow\left(1-x_{1}\right)x_{2}$, we find that\[
I_{4}^{4m}=\frac{3i}{32\pi}\int_{0}^{1}dx_{3}dx_{2}dx_{1}\left(1-x_{1}\right)^{-1/2}\left(1-x_{2}\right)\]
\[
\left\{ \left(1-x_{1}\right)\left(1-x_{2}\right)\left[x_{2}x_{3}m_{2}^{2}+x_{2}\left(1-x_{3}\right)t+\left(1-x_{2}\right)x_{3}\left(1-x_{3}\right)m_{3}^{2}\right]\right.\]
\begin{equation}
\left.+x_{1}\left[x_{2}m_{1}^{2}+\left(1-x_{2}\right)x_{3}s+\left(1-x_{2}\right)\left(1-x_{3}\right)m_{4}^{2}\right]+i\epsilon\right\} ^{-5/2}{\normalcolor .}\label{eq:i4mb}\end{equation}
For simplicity, we will set $\epsilon=0$ in the rest of the calculation.

Let's first evaluate the $x_{1}$ integral in eq \eqref{eq:i4mb}. To
do so, it is convenient to write the integral as follows:

\[
I_{4}^{4m}=\frac{3i}{32\pi}\int_{0}^{1}dx_{3}dx_{2}a\int_{0}^{1}dx_{1}\left(1-x_{1}\right)^{-1/2}\left(\left(c-b\right)x_{1}+b\right)\]
where

\[
a=1-x_{2}\]
\[
b=\left(1-x_{2}\right)\left[x_{2}x_{3}m_{2}^{2}+x_{2}\left(1-x_{3}\right)t+\left(1-x_{2}\right)x_{3}\left(1-x_{3}\right)m_{3}^{2}\right]\]
\[
c=x_{2}m_{1}^{2}+\left(1-x_{2}\right)x_{3}s+\left(1-x_{2}\right)\left(1-x_{3}\right)m_{4}^{2}{\normalcolor .}\]
Noting that

\[
\int_{0}^{1}dx_{1}\left(1-x_{1}\right)^{-1/2}\left(\left(c-b\right)x_{1}+b\right)=\frac{2\left(2b+c\right)}{3b^{3/2}c^{2}}{\normalcolor ,}\]
we obtain

\begin{equation}
I_{4}^{4m}=\frac{i}{16\pi}\int_{0}^{1}dx_{3}dx_{2}\frac{\left(1-x_{2}\right)\left[2\left(1-x_{2}\right)\left(\alpha x_{2}+\beta\left(1-x_{2}\right)\right)+m_{1}^{2}x_{2}+\gamma\left(1-x_{2}\right)\right]}{\left(1-x_{2}\right)^{3/2}\left(\alpha x_{2}+\beta\left(1-x_{2}\right)\right)^{3/2}\left(m_{1}^{2}x_{2}+\gamma\left(1-x_{2}\right)\right)^{2}}\label{eq:x2int}\end{equation}
where

\[
\alpha=x_{3}m_{2}^{2}+\left(1-x_{3}\right)t,\,\,\,\beta=x_{3}\left(1-x_{3}\right)m_{3}^{2},\,\,\,\gamma=x_{3}s+\left(1-x_{3}\right)m_{4}^{2}{\normalcolor .}\]
Performing the $x_{2}$ integral in eq \eqref{eq:x2int} then gives 

\begin{equation}
I_{4}^{4m}=-\frac{i}{8\pi}\int_{0}^{1}dx_{3}\left[\frac{\alpha\left(\beta-\gamma\right)+\beta\gamma}{\alpha\sqrt{\beta}\gamma\left(\alpha\gamma-m_{1}^{2}\beta\right)}+\frac{\alpha+\gamma-m_{1}^{2}}{m_{1}\left(m_{1}^{2}\beta-\alpha\gamma\right)^{3/2}}\tanh^{-1}\sqrt{1-\frac{\alpha\gamma}{m_{1}^{2}\beta}}\right]{\normalcolor .}\label{eq:x3int}\end{equation}

The integral in eq \eqref{eq:x3int} can be performed analytically using
Mathematica, for example. The final answer is presented in the end
of the attached Mathematica notebook. Although our final expression
for $I_{4}^{4m}$ in three dimensions is very complicated, we can
see two important features. First of all, it contains logarithms and
square roots of the kinematic variables. Second of all, it blows up
when any of the external masses approach zero, so the box integrals
with one or more massless external legs cannot be obtained from our final expression for the four-mass box integral. 

\subsection{3m box}

Reference \cite{Duplancic:2000sk} obtained an expression for the 3m box function in $d=4+2\epsilon_{IR}$ which is valid
for general $\epsilon_{IR}$. Replacing $\epsilon_{IR}$ with $-\frac{1}{2}+\tilde{\epsilon}$ in this expression
gives

\begin{equation}
I_{4}^{3m}=\frac{2i}{(4\pi)^{2}\left(st-m_{2}^{2}m_{4}^{4}\right)}\frac{\Gamma\left(\frac{3}{2}-\tilde{\epsilon}\right)}{\left(4\pi\mu^{2}\right)^{-\frac{1}{2}+\tilde{\epsilon}}}\left(P^{3m}+Q^{3m}\right).\label{eq:i3m}\end{equation}
$P^{3m} $ is given by
\[
P^{3m}=\frac{1}{2}\left(R\left(t,m_{2}^{2},m_{3}^{3}\right)+R\left(s,m_{4}^{2},m_{3}^{3}\right)\right)\]
where
\[
R\left(\alpha,\beta,m_{3}^{3}\right)=\frac{\Gamma\left(-\frac{1}{2}+\tilde{\epsilon}\right)}{\Gamma\left(-1+2\tilde{\epsilon}\right)}\left(1-\frac{\beta+i\epsilon}{\alpha+i\epsilon}\right)\sum_{n=0}^{\infty}\left(\frac{m_{3}^{2}+i\epsilon}{\alpha+i\epsilon}\right)^{n}\times\]
\[
\left[\left(-\alpha-\epsilon\right)^{-\frac{1}{2}+\tilde{\epsilon}}\frac{\Gamma\left(n-\frac{1}{2}+\tilde{\epsilon}\right)\Gamma\left(1+n\right)}{\Gamma\left(2+2n\right)}\mbox{}_{2}F_{1}\left(\frac{3}{2}+n-\tilde{\epsilon},1+n;2+2n,1-\frac{\beta+i\epsilon}{\alpha+i\epsilon}\right)\right.\]
\begin{equation}
\left.-\left(-m_{3}^{2}-\epsilon\right)^{-\frac{1}{2}+\tilde{\epsilon}}\frac{\Gamma\left(n-1+2\tilde{\epsilon}\right)\Gamma\left(\frac{1}{2}+n+\tilde{\epsilon}\right)}{\Gamma\left(1+2n+2\tilde{\epsilon}\right)}\mbox{}_{2}F_{1}\left(1+n,\frac{1}{2}+n+\tilde{\epsilon};1+2n+2\tilde{\epsilon},1-\frac{\beta+i\epsilon}{\alpha+i\epsilon}\right)\right]\label{eq:rsum}\end{equation}
assuming (without loss of generality) that $\left|m_{3}^{2}\right|<|\alpha|,|\beta|$. The functions $\mbox{}_{2}F_{1}$ are known as hypergeometric functions. 
Furthermore, $Q^{3m}$ is given by
\begin{equation}
Q^{3m}=\frac{\Gamma\left(-\frac{1}{2}+\tilde{\epsilon}\right)}{2\Gamma\left(\frac{3}{2}-\tilde{\epsilon}\right)\Gamma\left(-1+2\tilde{\epsilon}\right)}\int_{0}^{1}dz\frac{f}{z-z_{0}}\int_{0}^{1}dyy^{-1+2\tilde{\epsilon}}\left(1-y\right)^{\frac{1}{2}-\tilde{\epsilon}}\left[a\left(b+ay\right)^{-1}-c\left(d+cy\right)^{-1}\right]\label{eq:q3m}\end{equation}
where\[
a=-z(1-z)m_{3}^{2}+zs+(1-z)m_{4}^{2},\,\,\, b=-zs-(1-z)m_{4}^{2}-i\epsilon,\]
\[
c=-z(1-z)m_{3}^{2}+zm_{2}^{2}+(1-z)t,d=-zm_{2}^{2}-(1-z)t-i\epsilon,\]
\[
f=\left(-z(1-z)m_{3}^{2}-i\epsilon\right)^{-\frac{1}{2}+\tilde{\epsilon}}.\] 

Let's compute the above expressions in the limit that $\tilde{\epsilon}=0$. Notice that the prefactor of the sum in eq \eqref{eq:rsum} is $\mathcal{O}\left(\tilde{\epsilon}\right)$
due to the $\Gamma\left(-1+2\tilde{\epsilon}\right)$ in the denominator.
Assuming that $\beta\neq0$, the hypergeometric functions in eq \eqref{eq:rsum} can be computed
using the following integral formula:\begin{equation}
\mbox{}_{2}F_{1}\left(a,b;c,z\right)=\frac{\Gamma(c)}{\Gamma(b)\Gamma(c-b)}\int dt\frac{t^{b-1}\left(1-t\right)^{c-b-1}}{\left(1-tz\right)^{a}},\,\,\, z\neq1{\normalcolor .}\label{eq:hyperintegral}\end{equation}
Using this formula, we see that the hypergeometric functions in eq
\eqref{eq:rsum} are finite for all $n$. Furthermore, the prefactor
of the first hypergeometric function in the brackets is finite for
all $n$ and the prefactor of the second hypergeometric function in
the brackets is finite for $n>1$. Hence only $n=0,1$ terms in the
sum contribute. Using eq \eqref{eq:hyperintegral}, one finds that 
\[
\mbox{}_{2}F_{1}\left(1,\frac{1}{2}+\tilde{\epsilon};1+2\tilde{\epsilon},z\right)=\frac{1}{\sqrt{1-z}}+\mathcal{O}\left(\tilde{\epsilon}\right),\,\,\,\mbox{}_{2}F_{1}\left(2,\frac{3}{2}+\tilde{\epsilon};3+2\tilde{\epsilon},z\right)=\frac{4}{z^{2}}\left(\frac{2-z}{\sqrt{1-z}}-2\right)\]
where $z=1-\frac{\beta+i\epsilon}{\alpha+i\epsilon}$. Hence,\begin{equation}
R\left(\alpha,\beta,m_{3}^{2}\right)=2\pi\left(m_{3}^{2}-i\epsilon\right)^{-1/2}\left(\frac{z}{\sqrt{1-z}}+\left(\frac{m_{3}^{2}+i\epsilon}{\alpha+i\epsilon}\right)\frac{1}{z}\left(2-\frac{2-z}{\sqrt{1-z}}\right)\right)+\mathcal{O}\left(\tilde{\epsilon}\right){\normalcolor .}\label{eq:rbeta}\end{equation}

Now let's compute $Q^{3m}$ in eq \eqref{eq:q3m}. Let's focus on the $y$ integral, which has the form

\[
\int_{0}^{1}dyy^{-1+2\tilde{\epsilon}}\left(1-y\right)^{\frac{1}{2}-\tilde{\epsilon}}\left[a\left(b+cy\right)^{-1}-d\left(e+fy\right)^{-1}\right]{\normalcolor .}\]
This integral can be performed analytically and is given by 

\[
-\frac{y^{2\tilde{\epsilon}}}{2\tilde{\epsilon}\left(1+\tilde{\epsilon}\right)b^{2}d^{2}}\left\{ 2\tilde{\epsilon}a^{2}d^{2}yF_{1}\left(1+2\tilde{\epsilon},-\frac{1}{2}+\tilde{\epsilon},1,2+2\tilde{\epsilon},y,-\frac{ay}{b}\right)\right.\]

\begin{equation}
\left.\left.-2\tilde{\epsilon}c^{2}b^{2}yF_{1}\left(1+2\tilde{\epsilon},-\frac{1}{2}+\tilde{\epsilon},1,2+2\tilde{\epsilon},y,-\frac{cy}{d}\right)+\left(1+2\tilde{\epsilon}\right)bd(bc-ad)\mbox{}_{2}F_{1}\left(-\frac{1}{2}+\tilde{\epsilon},2\tilde{\epsilon};1+2\tilde{\epsilon},y\right)\right\} \right|_{y=0}^{y=1}
\label{alphaint}
\end{equation}
where $F_{1}$ is an Appell hypergeometric function. Noting that $F_{1}\left(1,-\frac{1}{2},1,2,0,0\right)=1$
and\[
F_{1}\left(1,-\frac{1}{2},1,2,1,\kappa\right)=\frac{2}{\kappa}\left(1-\sqrt{1/\kappa-1}\sin^{-1}\sqrt{\kappa}\right)\]
we see that the first two terms in the curly brackets in eq \eqref{alphaint} are $\mathcal{O}\left(\tilde{\epsilon}\right)$.
Furthermore, $\mbox{}_{2}F_{1}\left(-\frac{1}{2},0;1,0\right)=\mbox{}_{2}F_{1}\left(-\frac{1}{2},0;1,1\right)=1$,
so the third term in the curly brackets gives\[
\left.\mbox{}_{2}F_{1}\left(-\frac{1}{2}+\tilde{\epsilon},2\tilde{\epsilon};1+2\tilde{\epsilon},y\right)\right|_{y=0}^{y=1}=\mathcal{O}\left(\epsilon\right){\normalcolor .}\]
Noting that the prefactor in eq \eqref{eq:q3m} is $\mathcal{O}\left(\tilde{\epsilon}\right)$
due to the $\Gamma\left(-1+2\tilde{\epsilon}\right)$ in the denominator,
we find that $Q^{3m}=\mathcal{O}\left(\tilde{\epsilon}\right)$.

\subsection{2mh box}

The 2mh box integral can be obtained by setting $m_2^2=0$ in eqs \eqref{eq:i3m}, \eqref{eq:rsum}, and \eqref{eq:q3m}. Noting that $Q^{3m}$ in eq \eqref{eq:q3m} vanishes for all values of the mass parameters in 3d, we find that

\[
I_{4}^{2mh}=\frac{2i}{(4\pi)^{2}st}\frac{\Gamma\left(\frac{3}{2}-\tilde{\epsilon}\right)}{\left(4\pi\mu^{2}\right)^{-\frac{1}{2}+\tilde{\epsilon}}}P^{2mh}
\]
where
\begin{equation}
P^{2mh}=\frac{1}{2}\left(R\left(t,0,m_{3}^{3}\right)+R\left(s,m_{4}^{2},m_{3}^{3}\right)\right)\label{eq:p2mh}\end{equation}
and $R$ is given by eq \eqref{eq:rsum}. Notice that we cannot obtain $P^{2mh}$ in eq \eqref{eq:p2mh} by setting $\beta=0$ in eq \eqref{eq:rbeta}, i.e. the limit $m_2^2 \rightarrow 0$ is not smooth in eq \eqref{eq:rbeta}. Rather, we note that when $\beta=0$, the hypergeometric functions in eq \eqref{eq:rsum} are given by

\[
\mbox{}_{2}F_{1}\left(\frac{3}{2}+n-\tilde{\epsilon},1+n;2+2n,1\right)=\frac{\Gamma\left(2+2n\right)\Gamma\left(-\frac{1}{2}+\tilde{\epsilon}\right)}{\Gamma\left(1+n\right)\Gamma\left(\frac{1}{2}+n+\tilde{\epsilon}\right)}\]
\begin{equation}
\mbox{}_{2}F_{1}\left(\frac{3}{2}+n-\tilde{\epsilon},1+n;2+2n,1\right)=\frac{\Gamma\left(2+2n\right)\Gamma\left(-\frac{1}{2}+\tilde{\epsilon}\right)}{\Gamma\left(1+n\right)\Gamma\left(\frac{1}{2}+n+\tilde{\epsilon}\right)}{\normalcolor .}\label{eq:betaeqzero}\end{equation}
These expressions are finite for all $n$. Hence, as for the three-mass
box, only the $n=0,1$ terms in eq \eqref{eq:rsum} contribute. Using eq \eqref{eq:betaeqzero},
we therefore find that

\[
R\left(\alpha,0,m_{3}^{3}\right)=4\pi\left(m_{3}^{2}-i\epsilon\right)^{-1/2}\left(\frac{m_{3}^{2}+i\epsilon}{\alpha+i\epsilon}\right)+\mathcal{O}\left(\tilde{\epsilon}\right){\normalcolor .}\]

\subsection{2me, 1m, and 0m boxes}

The 2me box integral can be obtianed by setting $m_3^2=0$ in eqs \eqref{eq:i3m}, \eqref{eq:rsum}, and \eqref{eq:q3m}. Noting that $Q^{3m}$ in eq \eqref{eq:q3m} vanishes for all values of the mass parameters in 3d, we find that 
\begin{equation}
I_{4}^{2me}=\frac{2i}{(4\pi)^{2}\left(st-m_{2}^{2}m_{4}^{4}\right)}\frac{\Gamma\left(\frac{3}{2}-\tilde{\epsilon}\right)}{\left(4\pi\mu^{2}\right)^{\epsilon_{IR}}}P^{2me}\label{eq:2me}\end{equation}
Notice that we cannot obtain $P^{2me}$ by setting $m_3^2=0$ in eq \eqref{eq:rbeta}, i.e. the limit $m_3^2 \rightarrow 0$ is not smooth in eq \eqref{eq:rbeta}. On the other hand, reference \cite{Duplancic:2000sk} obtained an expression for $P^{2me}$ in $d=4+2\epsilon_{IR}$ which is valid for general $\epsilon_{IR}$. Replacing $\epsilon_{IR}$ with $-\frac{1}{2}+\tilde{\epsilon}$ in this expression
gives
\begin{equation}
P^{2me}=\frac{\Gamma^{2}\left(-\frac{1}{2}+\tilde{\epsilon}\right)}{\Gamma\left(2 \tilde{\epsilon}\right)}\left[\left(-s-i\epsilon\right)^{-\frac{1}{2}+\tilde{\epsilon}}+\left(-t-i\epsilon\right)^{-\frac{1}{2}+\tilde{\epsilon}}-\left(-m_{2}^{2}-i\epsilon\right)^{-\frac{1}{2}+\tilde{\epsilon}}-\left(-m_{4}^{2}-i\epsilon\right)^{-\frac{1}{2}+\tilde{\epsilon}}\right]\label{eq:P}\end{equation}
Expanding about $\tilde{\epsilon}=0$, we see that $P^{2me}$ is $\mathcal{O}(\tilde{\epsilon})$ due to the $\Gamma\left(2 \tilde{\epsilon}\right)$ appearing in the denominator of the prefactor. Hence, $I_4^{2me}$ vanishes in 3d using dimensional regularization. Since $I_{4}^{1m}$ and $I_{4}^{0m}$ can be obtained from  $I_{4}^{2me}$ by setting one or both of the external massess to zero, this implies that $I_{4}^{1m}$ and $I_{4}^{0m}$ also vanish.


\end{document}